\documentclass[10pt,journal,final,compsoc]{IEEEtran}
\ifCLASSOPTIONcompsoc
  \usepackage{cite}
\else
  \usepackage{cite}
\fi
\ifCLASSINFOpdf
\else
\fi
\newcommand{\eat}[1]{}
\usepackage{graphicx}
\usepackage{balance}  % for  \balance command ON LAST PAGE  (only there!)
\usepackage{booktabs}
\usepackage{tabularx}
\usepackage{amsfonts}
\usepackage{amsmath}
\usepackage{multirow}
\usepackage{algorithm}
\usepackage{algorithmic}
\usepackage{siunitx}

\usepackage{xcolor}
\usepackage{bm}

\usepackage{mathtools}
\usepackage{subfig}

\usepackage{bm}
\usepackage{balance}
\DeclarePairedDelimiterX{\infdivx}[2]{(}{)}{%
  #1\;\delimsize\|\;#2%
}
\newcommand{\infdiv}{D_{KL}\infdivx}

\def\Vec#1{{\boldsymbol{#1}}}
\def\Mat#1{{\boldsymbol{#1}}}
\usepackage[export]{adjustbox}

\usepackage{color}
\begin{document}
\title{Interpretable Signed Link Prediction with Signed Infomax Hyperbolic Graph}

\author{Yadan~Luo, Zi~Huang, Hongxu~Chen, Yang~Yang, Hongzhi Yin and Mahsa~Baktashmotlagh% <-this % stops a space
\IEEEcompsocitemizethanks{

\IEEEcompsocthanksitem Y. Luo, Z. Huang, M. Baktashmotlagh and H. Yin are with School of Information Technology and Electrical Engineering, The University of Queensland, Australia. E-mail: lyadanluol@gmail.com, huang@itee.uq.edu.au, m.baktashmotlagh@uq.edu.au, h.yin1@uq.edu.au.\protect
% \IEEEcompsocthanksitem K. Swersky ‬is with Google Research, Toronto, Canada. E-mail: kswersky@cs.toronto.edu.
\IEEEcompsocthanksitem H. Chen is with the School of Computer Science in University of Technology Sydney, Australia. E-mail: hongxu.chen@uts.edu.au.
\IEEEcompsocthanksitem Y. Yang is with the Center for Future Media and the School of Computer Science and Engineering,
University of Electronic Science and Technology of China, China. E-mail:  dlyyang@gmail.com.}
\thanks{Manuscript received November 25, 2020. Manuscript revised June 13, 2021.}}


\markboth{IEEE TRANSACTIONS ON KNOWLEDGE AND DATA ENGINEERING,~Vol.~14, No.~8, August~2020}%
{Shell \MakeLowercase{\textit{et al.}}: Bare Demo of IEEEtran.cls for Computer Society Journals}



\IEEEtitleabstractindextext{%
\begin{abstract}
Signed link prediction in social networks aims to reveal the underlying relationships (\textit{i.e.} links) among users (\textit{i.e.} nodes) given their existing positive and negative interactions observed. Most of the prior efforts are devoted to learning node embeddings with graph neural networks (GNNs), which preserve the signed network topology by message-passing along edges to facilitate the downstream link prediction task. Nevertheless, the existing graph-based approaches could hardly provide human-intelligible explanations for the following three questions: (1) which neighbors to aggregate, (2) which path to propagate along, and (3) which social theory to follow in the learning process. To answer the aforementioned questions, in this paper, we investigate how to reconcile the \textit{balance} and \textit{status} social rules with information theory and develop a unified framework, termed as Signed Infomax Hyperbolic Graph (\textbf{SIHG}). By maximizing the mutual information between edge polarities and node embeddings, one can identify the most representative neighboring nodes that support the inference of edge sign. Different from existing GNNs that could only group features of friends in the subspace, the proposed SIHG incorporates the signed attention module, which is also capable of pushing hostile users far away from each other to preserve the geometry of antagonism. The polarity of the learned edge attention maps, in turn, provides interpretations of the social theories used in each aggregation. In order to model high-order user relations and complex hierarchies, the node embeddings are projected and measured in a hyperbolic space with a lower distortion. Extensive experiments on four signed network benchmarks demonstrate that the proposed SIHG framework significantly outperforms the state-of-the-arts in signed link prediction.
\end{abstract}

}

\maketitle

\IEEEdisplaynontitleabstractindextext

% \fi
\IEEEpeerreviewmaketitle

\IEEEraisesectionheading{\section{Introduction}\label{sec:introduction}}

% 1. Motivation (Hashing -> Deep Hashing : (1)Symmetric (2)Asymmetric)

\IEEEPARstart{U}{nderstanding}  social interactions on the Web is critical for a broad set of tasks, such as community detection~\cite{community1,community2,community3}, personalized recommendation \cite{personalized, personalized1}, fake account detection \cite{www} and event prediction~\cite{eventpred}. By giving thumbs up, following and subscribing, users expose their positive preferences, support and approval for others who share the same opinions. Users also link to signify disapproval, disagreement, or distrust of others with negative response such as blocking. In such a signed social network, users or entities of interest are generally represented as nodes, and the mutual interactions are modeled as edges (or links) with signs.

While important, the interplay of positive and negative relations poses a great challenge to the vast majority of conventional online social network research that only considers observed connections as positive links, which refers to \textit{unsigned} networks. Dedicated network embedding methods for unsigned networks \cite{unsigned,unsigned1,unsigned2,unsigned3,unsigned4,unsigned5,unsigned6} developed in the past, have exploited the fact that the node embeddings are highly correlated with the link structure of the network. These algorithms, therefore, predominantly focus on node representations and force the connected nodes to have similar latent features to preserve both local connectivity and the high-order proximity among nodes in a graph. By using the low-dimensional node vectors, traditional machine learning methods can be applied to predict the connectivity between an arbitrary pair of nodes. However, for the real-life networks which consist of both positive and negative links (\textit{i.e.} signed networks), the existing unsigned link prediction techniques are not directly applicable by virtue of lacking a specific mechanism to deal with the negative connections.

% The extracted motifs then provide different aggregation paths for each node, taking the various neighborhoods into consideration.

% simply using attention is not clear (massive neighbors); lack a golden rule to unify different theories -> Mutual Information Maximization to guide
% dilemma: defining local motifs in sub-branches (heavy); automatically choose (could hardly preserve) -> Hyperbolic space
% existing graph-based net cannot push away negative pairs -> (-1, 1) attention score
\begin{figure}[t]
    \centering
    \includegraphics[width=1\linewidth]{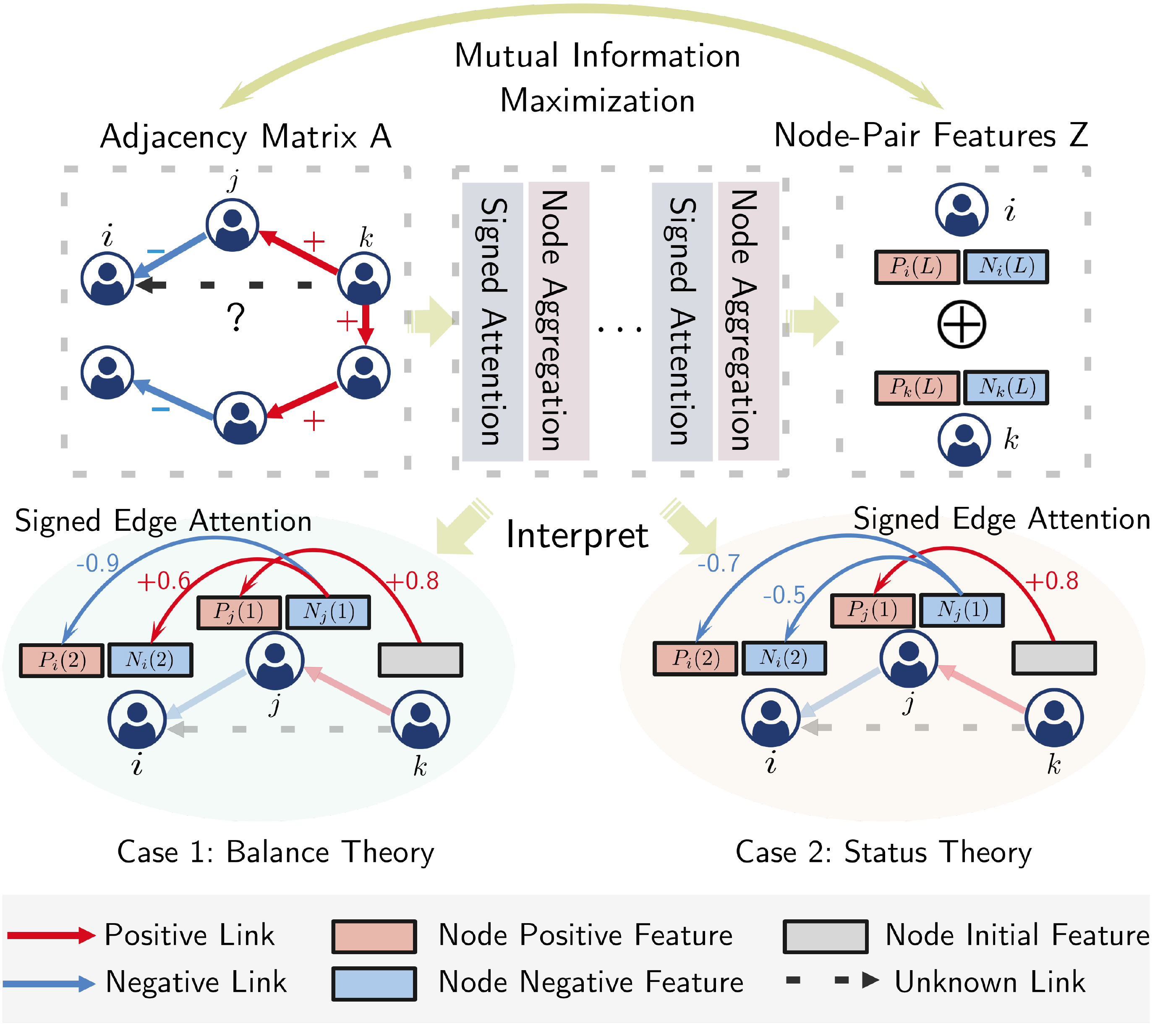}
    \caption{An illustration of the interpretability of the proposed model for signed social networks. The learned signed attention maps are optimized by maximizing the mutual information between the node representations and the given edge polarity, which provide cues for discovering the underlying social theories.}
    \label{fig:demo}
\end{figure}
% Signed 
% Lately, a line of attempts has investigated signed node embeddings by leveraging signed Laplacian analysis~\cite{clustering,SSE}, matrix factorization~\cite{MF}, probablistic strategies~\cite{SNE}, and random walk~\cite{SIDE,SIGNet} to capture the structural properties of \textit{signed} networks. Another stream of work applies deep metric learning~\cite{SiNE} by ranking the distances of positively- and negatively- linked node pairs. 

% Graph-based Signed & two theories
With a rapid development of graph neural networks (GNNs), recent works of SGCN~\cite{SGCN} and SNEA~\cite{SNEA} have been focusing on re-designing deep graph models to work with \textit{signed} and undirected graph structures, where node features are recursively aggregated with adjacent nodes along signed edges. The core idea is based on one of the well-known social theories, the \textbf{structural balance theory}~\cite{balancetheory}. It roughly implies that attitudes of a user can change based on the assumption that a user's \textit{friends of friends} can be deduced as friends, while \textit{enemies of friends} can be considered as \textit{enemies}. Nevertheless, as the balance rule is only tenable under certain circumstances, the \textbf{status theory}~\cite{guha, slashdot, jure} is developed, which states that the signed links can be inferred by comparing the users' social status.

As modern social networks are complex and evolve quickly, user preferences or opinions remain uncertain and unstable. Before reaching the convergence, the underlying signed links cannot be explicitly predicted by either social theory. For instance, as shown in Figure \ref{fig:demo}, given the positive link from user $k$ to user $j$ and the negative link from user $j$ to user $i$, user $k$ will be predicted as an enemy of user $i$ based on the \textbf{balance theory}, while as a friend if it is based on the \textbf{status theory}. Thereby, in our work, we aim to better comprehend the complex interplay between these two social theories and jointly accommodate them in one unified framework to provide reasonable explanations of the user interactions at the current stage.

\begin{figure}[t]
    \centering
    \includegraphics[width=0.49\linewidth]{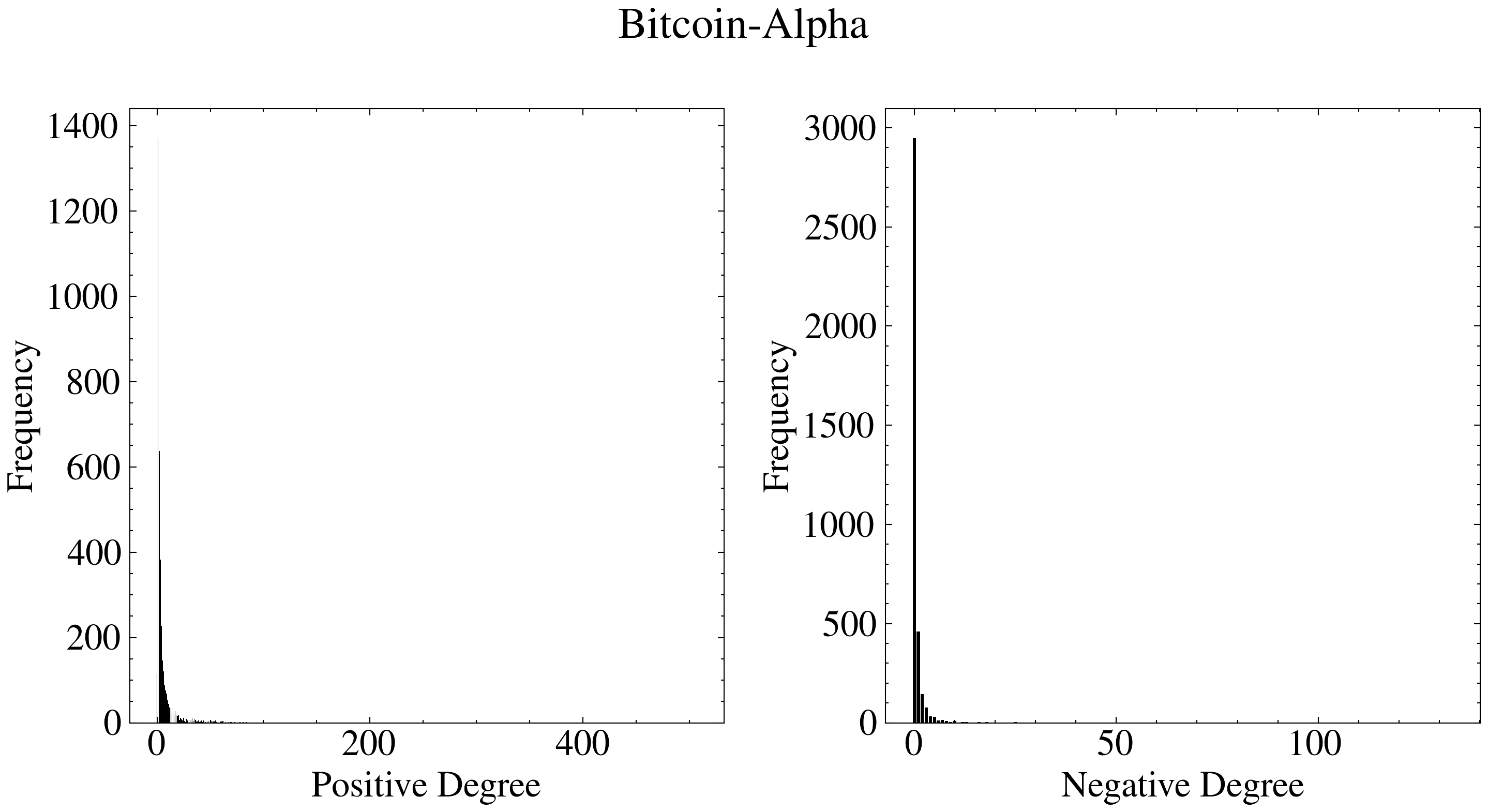}
    \includegraphics[width=0.49\linewidth]{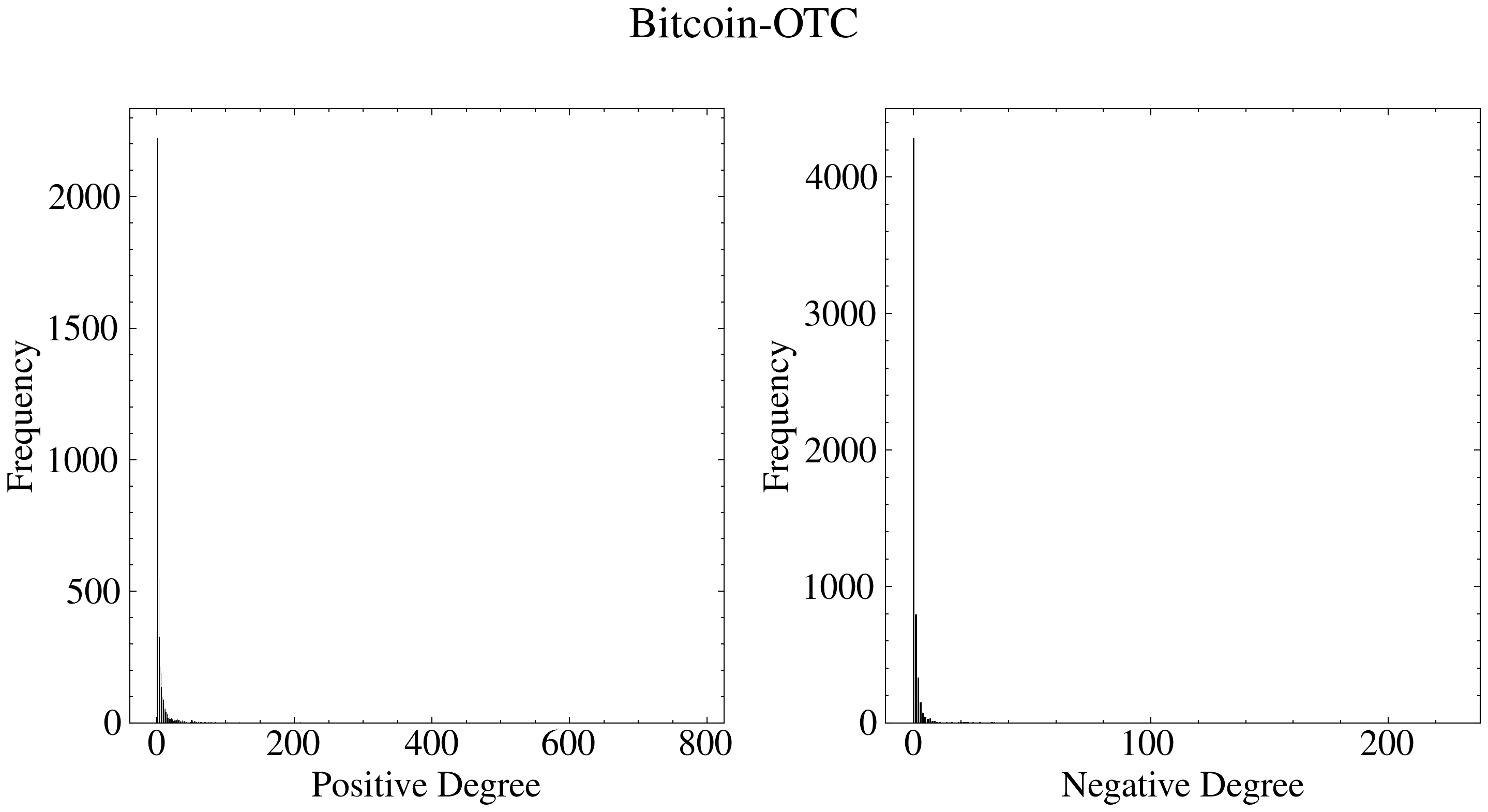}\\
    \includegraphics[width=0.49\linewidth]{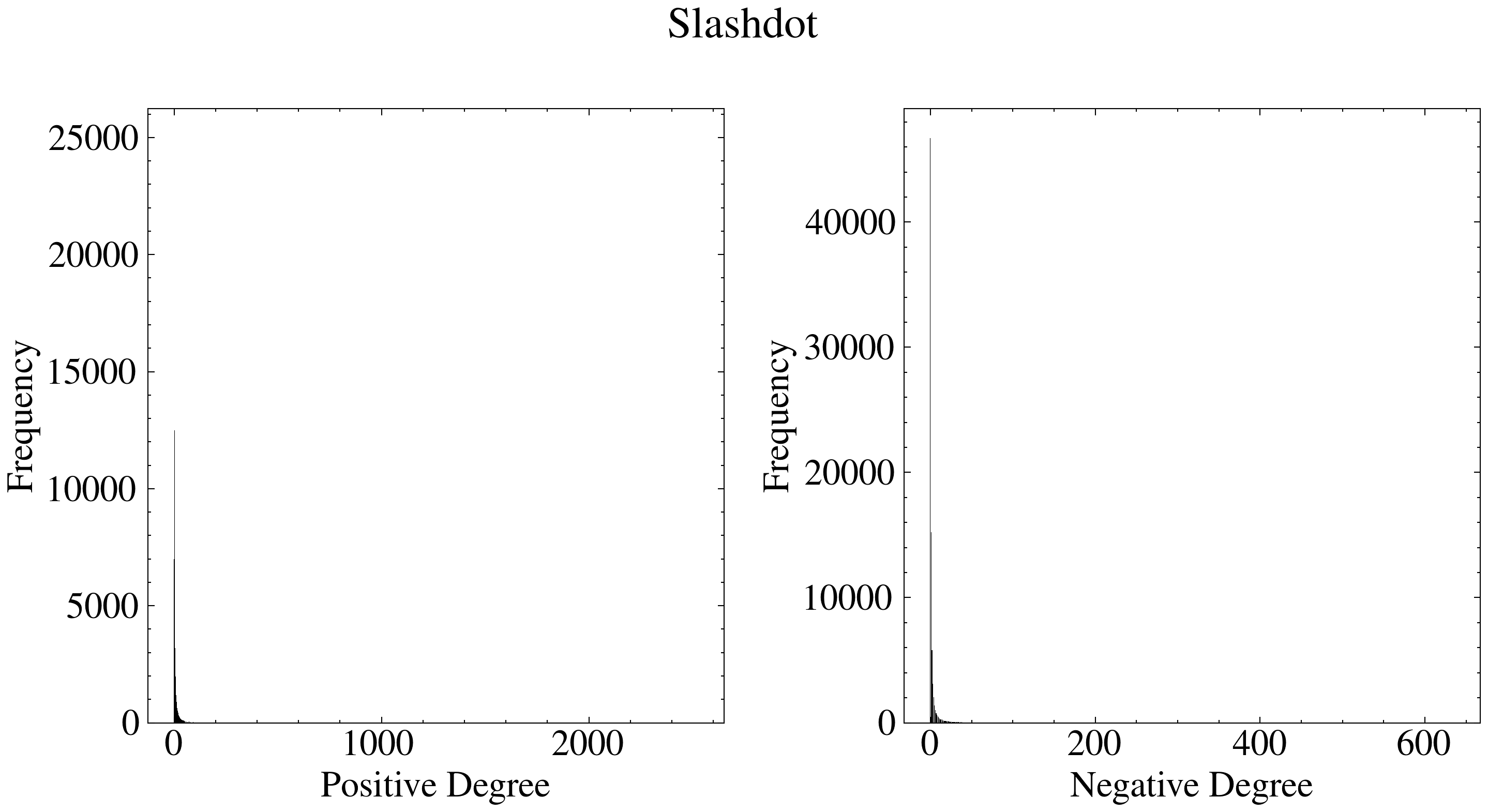}
    \includegraphics[width=0.49\linewidth]{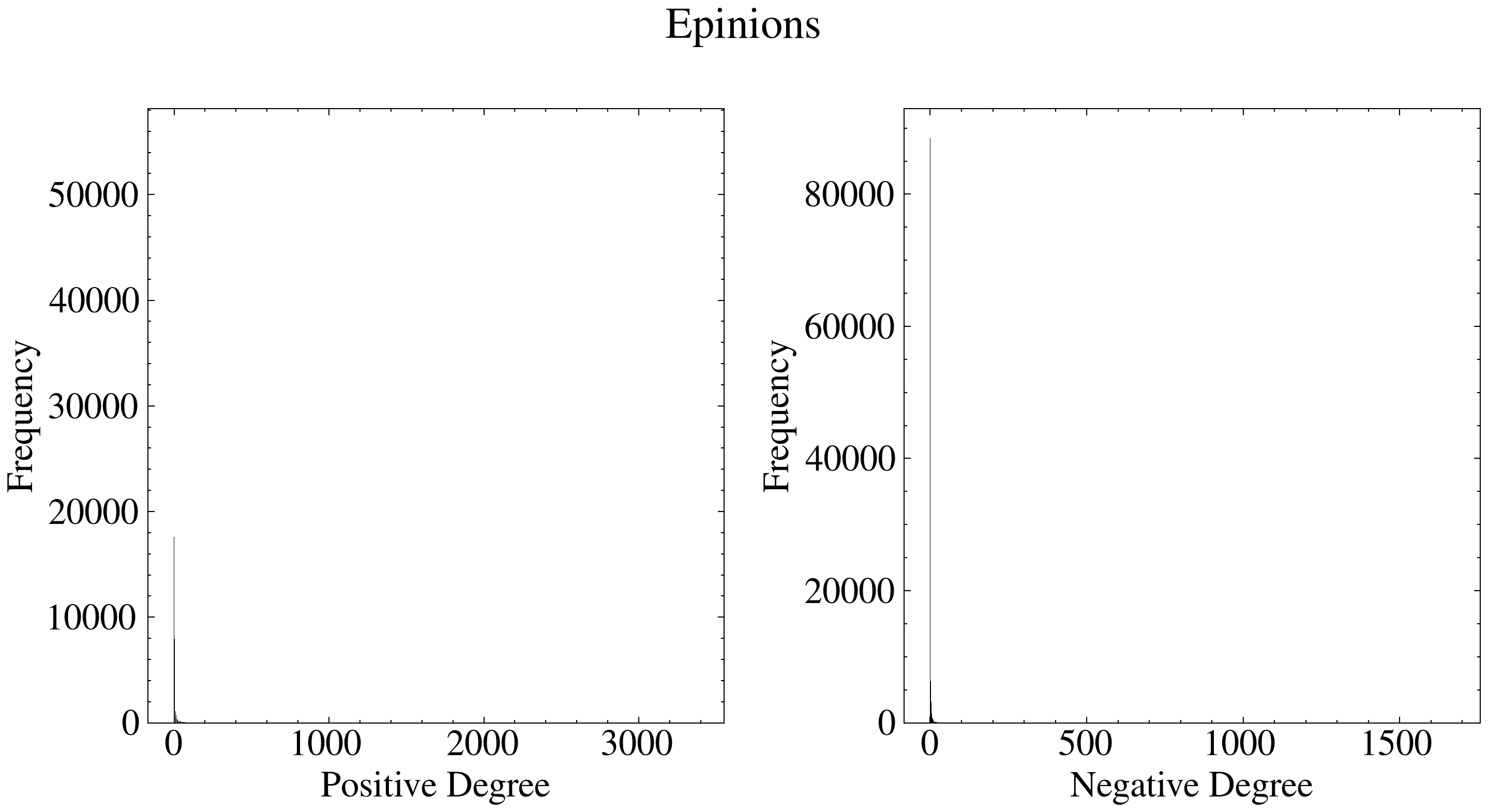}
    \caption{\color{black}Positive and negative distributions in signed social networks. The power-law distributions exhibited implies the underlying node hierarchy is complex, which can not be well captured by Euclidean embeddings.}
    \label{fig:degree1}
\end{figure}

Even though one of recent work~\cite{SiGAT} that attempts to combine two social theories by manually defining 38 different types of local network motifs for guiding aggregation, it is still non-trivial to propose such a unified framework due to the following three challenges:
\begin{itemize}
    \item The existing graph-based models generally lack interpretability in a sense that they do not easily allow for human-intelligible explanations of the following questions: for each target node-pair, (1) which neighboring nodes are decision-critical? (2) how to design the aggregation path for multi-hop neighbors? and (3) which social theory can be referred to? While the first two questions can be answered by inferring edge attention scores~\cite{SNEA,SiGAT}, the last one has not yet been investigated. 
    
    \item On the one hand, manually defining local motifs for aggregation is tedious and time-consuming, especially in the presence of massive number of adjacent neighbors. On the other hand, automatically embedding complex hierarchical neighborhood structures in GCNs or GAT can incur a large distortion~\cite{hgcn,hnn}. \color{black}As depicted in Figure \ref{fig:degree1}, the positive and negative connections in signed social networks at a large scale typically exhibit power-law distributions, which implies the underlying node hierarchy is complex. \color{black}Such a dilemma forces common graph-based models to trade off between human-labors and prediction accuracy.
    
    \item Existing graph models do not define an inverse operation of grouping features, which means that features of negative pairs could hardly be pushed away from each other in the embedding space.
\end{itemize}

\noindent \textbf{Core idea:} To address the first challenge of endowing the models with interpretability, we demonstrate our core idea in Figure \ref{fig:demo}. Inspired by recent advances in information theory, our strategy is to learn the optimal signed edge attention scores by maximizing the mutual information between the adjacency matrix of the observed social networks and
the concatenated representations of the node pairs. By doing so, (1) the target nodes can automatically identify the most representative positive or negative neighbors multi-hops away that could support the prediction of the given edge polarity; (2) the explanations for each aggregation can be discovered by comparing the signs of the learned attention scores. For example, \color{black}{as shown in the Case 1 of Figure \ref{fig:demo}, each node feature is comprised of a positive part (e.g., $P_i(1)$) and a negative part (e.g., $N_i(1)$)}, where they represent the information of the positive and negative neighborhood and $1$ indicates the first aggregation layer. \color{black}When aggregating the user $j$'s negative features $N_j(1)$ at the first graph layer to the user $i$'s negative features $N_i(2)$ at the second layer, one can identify the balance theory is applied if \color{black}the attention score $w_{ij}^{-(2)}$ is positive (e.g., $w_{ij}^{-(2)} = +0.6$); otherwise, the status rule is leveraged in this aggregation (e.g., $w_{ij}^{-(2)} = -0.6$). \color{black}

Specifically, in this paper, we propose a novel Signed Infomax Hyperbolic Graph (\textbf{SIHG}) network to reconcile \textit{balance} and \textit{status} theories in the task of signed link prediction. In order to fully inherit the rich hierarchical information in signed graphs, we generalize graph neural network to operate on a non-Euclidean space, where features from positive neighbors and negative neighbors are separately aggregated in different hyperbolic manifolds. To jointly realize the inclusion of features for friends and exclusion for foes in the embedding space, a signed attention mechanism is further incorporated, where positively linked nodes are mapped to close-by points whereas negatively linked nodes are transformed far from each other in hyperbolic manifolds. In line with the mutual information estimation, the learned attention scores provide interpretable explanations of social theories. As the proposed SIHG is agnostic to the choice of hyperbolic models, two hyperbolic models are testified in our framework. Extensive experiments conducted on four signed social network benchmarks evidence the superiority of the derived SIHG over the state-of-the-art approaches. In summary, our contribution is four-fold.

\begin{itemize}
    \item We introduce a new Signed Infomax Hyperbolic Graph (SIHG) framework for signed link prediction, which unifies two social theories and provides human-intelligible interpretations by maximizing mutual information (addressing challenge 1).
    \item The projected hyperbolic space fully exploits the topology of users' positive and negative neighborhoods and learns the respective geometrical representations. Two hyperbolic models of Hyperboloid and Poincaré Ball are testified in our framework (addressing challenge 2).
    \item By incorporating the mutual information estimation, the derived signed attention module not only automatically learns the aggregation paths, but also pushes the hostile user nodes far away from each other (addressing challenge 3).
    \item We have demonstrated the effectiveness and interpretability of the proposed SIHG through extensive quantitative experiments and qualitative visualizations on four large-scale signed social network datasets. Source code\footnote{\color{black}https://github.com/Luoyadan/SIHG} is provided for reference.
\end{itemize}

The rest of the paper is organized as follows. Section 2 introduces the mathematical definition of the signed link prediction task and theoretical foundations of signed social networks, followed by the details of the proposed SIHG model. The experimental comparisons with state-of-the-art, ablation study and visualizations are highlighted in Section 4. Section 5 presents a brief review of recent advances in signed link prediction and mutual information estimation. We conclude in Section 6.
\begin{figure}[t]
    \centering
    \includegraphics[width=1\linewidth]{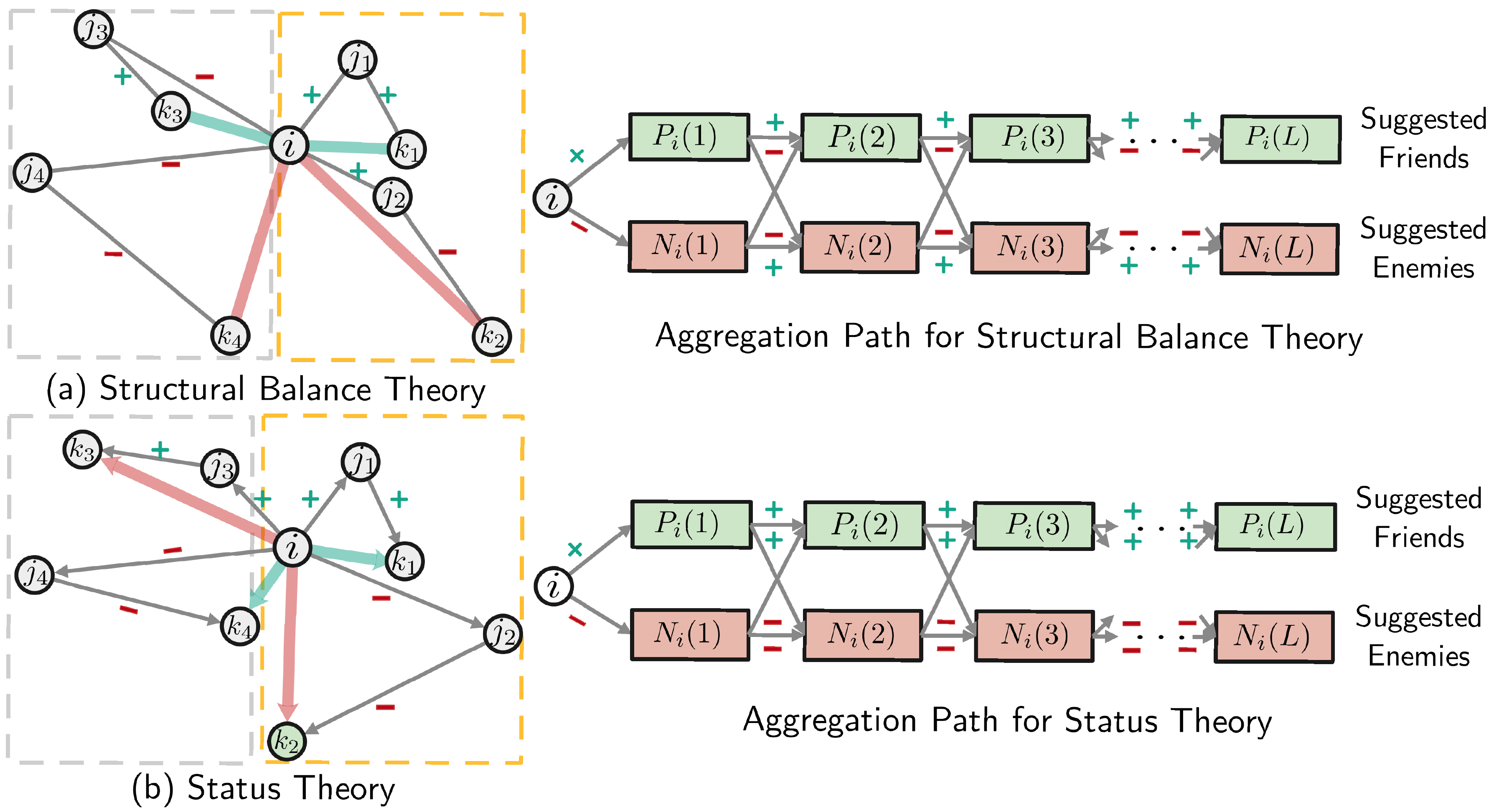}
    \caption{An illustration of the (a) structural balance and (b) status theories, and the suggested aggregation paths. \color{black} The gray line with a `+' symbol refers to a positive connection, while the one with a `-' symbol refers to a negative connection. Green edges and red edges in the left figure indicates the inferred positive and negative relations, respectively. The gray boxes contain the triangles that do not satisfy the theory and yellow boxes include }
    \label{fig:theory}\vspace{-3ex}
\end{figure}

\begin{figure*}[t]
    \centering
    \includegraphics[width=0.99\linewidth]{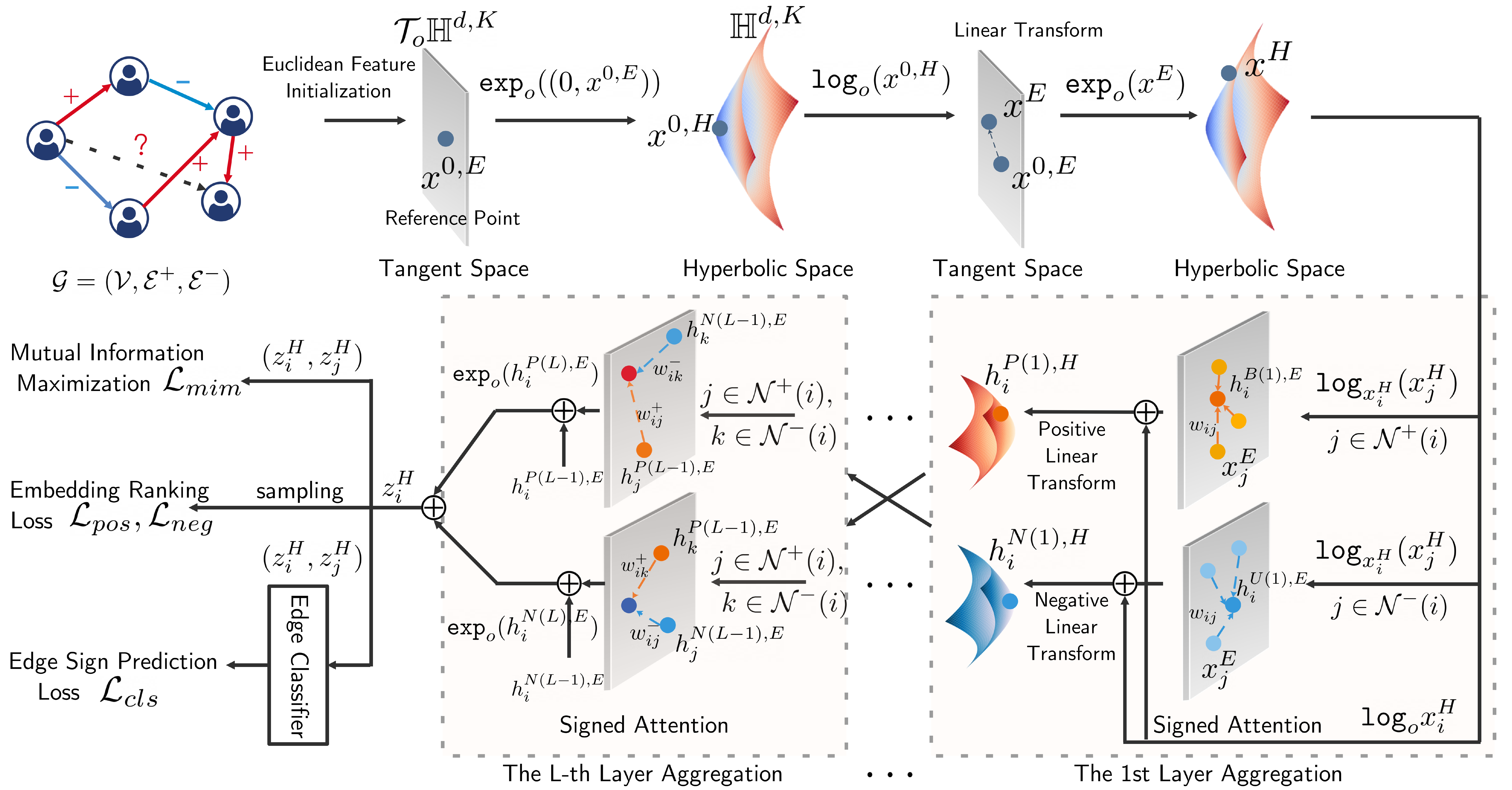}
    \caption{An overview of the proposed Signed Infomax Hyperbolic Graph (SIHG) architecture with $L$ graph layers.}
    \label{fig:flowchart}\vspace{-3ex}
\end{figure*}

\begin{table}[h]
    \centering
    \caption{\color{black}Summary of frequently used notation.}
    \resizebox{1\linewidth}{!}{% 
    \begin{tabular}{>{\color{black}}l >{\color{black}}l}
    \toprule
    Symbol &Description\\
    \midrule
      {\textbf{Input Graph}: } \\
      $\mathcal{G}$ & Signed social network\\ 
      $\mathcal{V}$ & Vertices\\
      $\mathcal{E}^+$ & Positive edge sets\\
      $\mathcal{E}^-$ & Negative edge sets\\
      $\Mat{A}$ &Adjacency matrix\\
      $\Vec{v}_i$ &The $i$-th vertex\\
      $x_i$\\
      \midrule
      {\textbf{Hyperbolic}: } \\
      $c$ &The curvature of hyperbolic space \\
      $\Vec{o}$ &The origin in $\mathbb{H}^{d,K'}$\\
      $\mathcal{T}_{\Vec{o}}\mathbb{H}^{d,K}$ &The tangent space centered at point $\Vec{o}$\\
      $\mathcal{T}_{\Vec{x}}\mathbb{H}^{d,K}$ &The tangent space centered at point $\Vec{x}$\\
      $\mathbb{R}^d$ &Euclidean space with dimension $d$\\
      \midrule
      {\textbf{Operations}: } \\
      $P_{o\rightarrow x}^K(\cdot)$ &Parallel transport from $\mathcal{T}_{\Vec{o}}\mathbb{H}^{d,K}$ to $\mathcal{T}_x\mathbb{H}^{d,K}$\\
      $\otimes^H$ &Multiplication in hyperbolic space\\
      $\oplus^H$ &Addition in hyperbolic space\\
      $\texttt{exp}_{\Vec{o}}$ and $\texttt{exp}_{\Vec{x}}$ &exponential map\\
      $\texttt{exp}_{\Vec{o}}$ and $\texttt{log}_{\Vec{x}}$ &logarithmic map\\
      $\texttt{Dist}(\Vec{x}, \Vec{y})$ &Intrinsic distance between $\Vec{x}$ and $\Vec{y}$ in $\mathbb{H}^{d,K}$\\
      $\|\Vec{v}\|_L$ &The norm of $\Vec{v}$ in $\mathcal{T}_x\mathbb{H}^{d,K}$\\
      \midrule
      {\textbf{Network}: } \\
      $\mathcal{G}_P$ &Positive branch of SIHG\\
      $\mathcal{G}_N$ &Negative branch of SIHG\\
      $\mathcal{G}_P^{(l),H}$ &The $l$-th layer of positive branch\\
      $\mathcal{G}_N^{(l),H}$ &The $l$-th layer of negative branch\\
      $\mathcal{V}^{P(l),H}$ &The vertices at the $l$-th layer in the positive branch\\
      $\mathcal{V}^{N(l),H}$ &The vertices at the $l$-th layer in the negative branch\\
      $\mathcal{N}^+(i)$ &The positive neighbor sets of the $i$-th vertice\\
      $\mathcal{N}^-(i)$ &The negative neighbor sets of the $i$-th vertice\\
      $h_i^{P(l), H}$ &The $i$-th vertex's feature at the $l$-th layer in the positive branch\\
      $h_i^{N(l), H}$ &The $i$-th vertex's feature at the $l$-th layer in the negative branch\\
      $z_i^{H}$ &The final representation of the $i$-th vertex\\
      $p_{ij}$ &The edge prediction between the $i$-th vertex and the $j$-th vertex\\
      \bottomrule
     \end{tabular}
     }\vspace{-2ex}
\end{table}

\section{Preliminaries}
\subsection{Problem Definition}
 A signed social network can either be modeled as a directed or undirected graph $\mathcal{G} = (\mathcal{V}, \mathcal{E}^+, \mathcal{E}^-)$ with a sign on each edge $e_{ij}\in\mathcal{E}^+\cup\mathcal{E}^-$, where the vertices $\mathcal{V} = \{\Vec{v}_i\}_{i=1}^{N}$ represent a set of $N$ users. The initial feature for each node is denoted as $\Vec{v}_i\in\mathcal{V}$ with $\Vec{x}_i\in\mathbb{R}^d$, where $d$ is the dimension of node embeddings. $\Mat{A}\in\mathbb{R}^{N\times N}$ is the adjacency matrix of the signed network, where $\Mat{A}_{ij} = 1$, if there is a positive link from $\Vec{v}_i$ to $\Vec{v}_j$, and $\Mat{A}_{ij} = 0$ for a negative link. For all $\Mat{A}_{ij}$ which are equal to 0.5, there is no directed edge from $\Vec{v}_i$ to $\Vec{v}_j$. Positive edges $\mathcal{E}^+ = \{e_{ij}|\Mat{A}_{ij}=1\}_{i,j=1}^{N}$, and negative edges $\mathcal{E}^- = \{e_{ij}|\Mat{A}_{ij}=0\}_{i,j=1}^N$ are partially observed at the training stage. The signed link prediction problem which we address here is to predict if a pair of nodes in $\mathcal{V}$ will remain disconnected or will be connected by a positive or negative link. 

 \subsection{Balance and Status Theories}
We start by introducing the two fundamental social-psychological theories, which are illustrated in Figure~\ref{fig:theory}. 

\noindent\textbf{Structural Balance Theory}~\cite{jure}: is based on the common principles that ``the friend of my friend is my friend'' and ``the foe of my friend is my foe''. For instance, in Figure~\ref{fig:theory}(a), if node $\Vec{v}_{k_1}$ forms a triad with the edge $e_{ij_1}$, the triangle on $(\Vec{v}_i, \Vec{v}_{j_1}, \Vec{v}_{k_1})$ should have an odd number of positive signs regardless of edge direction. Those triads that obey structural balance theory refer to \textit{balanced triangles}, such as $(\Vec{v}_i, \Vec{v}_{j_1}, \Vec{v}_{k_1})$ and $(\Vec{v}_i, \Vec{v}_{j_2}, \Vec{v}_{k_2})$ shown in the yellow dotted box. Otherwise, they are called as unbalanced triangles as highlighted in the grey dotted box. 

\noindent\textcolor{black}{\textbf{Status Theory}: Alternatively, another theory is developed based on the notion of status~\cite{guha, slashdot}, as shown in Figure \ref{fig:theory}(b). The status may indicate the relative prestige, ranking, or reputation. The status theory posits that, in a positive edge $e_{ij}$, user $i$ regards user $j$ as having a higher status, while in a negative edge $e_{ij}$, user $i$ regards user $j$ as having a lower status. By assuming that all nodes follow the status ordering, the edge sign should flip if its direction is flipped. }

\noindent \textcolor{black}{\textbf{Comparison}. Based on different social theories, the latent representation of each node $i$ in a signed graph will be updated along various aggregation paths, which are demonstrated in Figure~\ref{fig:theory}. For instance, according to balance theory, the first-order positive neighbors (e.g., $\Vec{v}_{j_2}$) of node $\Vec{v}_i$ and the second-order negative neighbors (e.g., $\Vec{v}_{k_2}$) will be grouped in $\Mat{P}_i(1)$ and $\Mat{N}_i(2)$, respectively. To this end, the edge polarity between $\Vec{v}_i$ and $\Vec{v}_{k_2}$ can be inferred by measuring the similarity of the respective node embeddings. }

\noindent\textbf{Discussion.} By comparing each of four typical types of signed triangles, the following two observations can be drawn:

\begin{enumerate}
    \item The edge sign cannot be simply inferred by either of the theories, due to the conflicts in some cases. For example, the triangle $(i, j_2, k_2)$ that satisfies status theory are not balanced. 
    \item For the same signed social network, the learned node embeddings can still vary significantly according to different theories. For example, with negative links $e_{ij_2}$ and $e_{j_2k_2}$, the representation of node $\Vec{v}_{k_2}$ should be far from the feature of node $\Vec{v}_i$ in status theory, which is opposite to the situation in balance theory.
\end{enumerate}
\noindent Motivated by the observations mentioned above, we aim to derive a unified framework that dynamically chooses a proper path to aggregate node embeddings and predict edge polarities. The core principle behind the framework is to infer the missing signs of edges that provides interpretable explanations for the given edge labels, which can be achieved by maximizing the mutual information in the signed social networks. 
\vspace{-3ex}
\section{Methodology}
In this section, we go through the details of the proposed Signed Infomax Hyperbolic Graph (\textbf{SIHG}) framework as illustrated in Figure \ref{fig:flowchart}. \color{black}In order to embed the positive and negative neighbors' information hierarchically and structurally, we firstly split the node representations into two part and construct two $L$-layer hyperbolic graph sub-networks, \textit{i.e.}, $\mathcal{G}_P = \{\mathcal{G}_P^{(l),H}=(\mathcal{V}^{P(l),H}, \mathcal{E}^+, \mathcal{E}^-)\}_{l=1}^L$ and $\mathcal{G}_N = \{\mathcal{G}_N^{(l),H}= (\mathcal{V}^{N(l),H}, \mathcal{E}^+, \mathcal{E}^-)\}_{l=1}^L$ for passing messages of positive and negative neighbors $L$-hops away, respectively. \color{black}The superscript $^H$ denotes hyperbolic embedding and the superscripts $^P$ and $^N$ indicate the positive and negative semantics, respectively. Each positive node $\Vec{v}_i^{P(l),H}\in\mathcal{V}^{P(l),H}$ and each negative node $\Vec{v}_i^{N(l),H}\in\mathcal{V}^{N(l),H}$ at the $l$-th layer are associated with hidden features of $\Vec{h}_i^{P(l),H}\in\mathbb{H}^{d,K}$ and $\Vec{h}_i^{N(l),H}\in\mathbb{H}^{d,K}$, where $\mathbb{H}^{d,K}$ is the hyperbolic manifold in $d$ dimensions with constant negative curvature $c=-1/K (K>0)$. As the input node feature $x_i^{0,E}\in\mathbb{R}^d$ is in a Euclidean space (denoted with the superscript $^E$), we first transform it to the hyperbolic space via the $\texttt{exp}_{\Vec{o}}$ map (Section \ref{sec:transform}), and then leverage the signed attention strategy to aggregate the features from neighborhoods (Section \ref{sec:agg}). The aggregation is guided by maximizing the mutual information between the concatenated node features and edge signs, as discussed in Section \ref{sec:mim}. To this end, two fundamental social theories can be seamlessly reconciled and interpreted by the learned attention maps.

\subsection{Hyperbolic Transformation}\label{sec:transform}
Before formalizing the hyperbolic transformations for input Euclidean node features, we start by outlining a relationship between the Euclidean space and Hyperbolic space. A hyperbolic space is a homogeneous space that has a constant negative curvature, which is distinguished from Euclidean spaces with zero curvature ($c=0$). While the tangent spaces of the points on hyperbolic manifold are isometric to $\mathbb{R}^d$, the Euclidean transformation can be performed in $\mathcal{T}_{\Vec{o}}\mathbb{H}^{d,K}$, where $\Vec{o}:=\{\sqrt{K},0,\ldots,0\}\in\mathbb{H}^{d,K}$ is the origin in $\mathbb{H}^{d,K}$. Motivated by this, we regard $(0, \Vec{x}^{0,E})$ as a point in the tangent space $\mathcal{T}_{\Vec{o}}\mathbb{H}^{d,K}$ and project it to hyperbolic space by $\texttt{exp}_{\Vec{o}}$ map, where $\Vec{o}$ serves as a reference point to : 
\begin{eqnarray}
    \Vec{x}^{0, H} = \texttt{exp}_{\Vec{o}}(0, \Vec{x}^{0, E}),
\end{eqnarray}
with $\Vec{x}^{0,H}$ the corresponding point in $\mathbb{H}^{d,K}$.
For linear transformation of a point in a Hyperbolic space, we define the following operations:

\begin{eqnarray}
   \begin{split}
        &\Vec{x}^H = \Mat{W} \otimes^H \Vec{x}^{0, H} \oplus^H \Vec{b},\\
        & \Mat{W}\otimes^H \Vec{x} := \texttt{exp}_{\Vec{o}}(\Mat{W} \texttt{log}_{\Vec{o}}(\Vec{x})),\\
        &\Vec{x} \oplus^H \Vec{b} := \texttt{exp}_{\Vec{x}}(P_{o\rightarrow x}^K(\Vec{b})),
    \end{split}
\end{eqnarray}
where $\Mat{W}\in\mathbb{R}^{d\times d}$ is a learnable weight matrix, and $\Vec{b}\in\mathbb{R}^d$ is a bias. The $P_{o\rightarrow x}^K(\cdot)$ is the parallel transport from $\mathcal{T}_{\Vec{o}}\mathbb{H}^{d,K}$ to $\mathcal{T}_x\mathbb{H}^{d,K}$. There are several important models of hyperbolic space such as the Klein model, Hyperboloid model, and Poincaré ball model, in which the differentiable operations $\texttt{log}$ and $\texttt{exp}$ can be implemented differently. Any two of the aforementioned models can be related by a transformation that preserves all the geometrical properties of the space, including isometry. Without lose of generality, here, we introduce the following two hyperbolic models in our framework:

\subsubsection{Hyperboloid Model}
The hyperboloid model, also called as Lorentz model, is defined as $\mathbb{H}^{d,K}:=\{\Vec{x}\in\mathbb{R}^{d+1}: \langle \Vec{x}, \Vec{x}\rangle_{L}=-K, x_0>0\}$, where $\langle\cdot,\cdot\rangle_{L}:\mathbb{R}^{d+1}\times\mathbb{R}^{d+1}\rightarrow \mathbb{R}$ indicating the Minkowski inner product, specifically $\langle \Vec{x}, \Vec{y}\rangle_L := -x_0y_0 + x_1y_1 + \ldots +x_dy_d$. The tangent space centered at point x is $\mathcal{T}_x\mathbb{H}^{d,K}:=\{\Vec{v}\in\mathbb{R}^{d+1}: \langle \Vec{v}, \Vec{x}\rangle_L=0\}$. The mapping between tangent space and hyperbolic space is through exponential and logarithmic maps, respectively. The exponential map $\texttt{exp}_{\Vec{x}}: \mathcal{T}_{\Vec{x}} \mathbb{
H}^{d,K}\rightarrow \mathbb{H}^{d,K}$, and the logarithmic map $\texttt{log}_{\Vec{x}}: \mathbb{H}^{d,K}\rightarrow  \mathcal{T}_{\Vec{x}}\mathbb{H}^{d,K}$ of the hyperboloid model are given by:

\begin{equation}
    \begin{split}
        &\texttt{exp}_{\Vec{x}}(\Vec{v}) = \text{cosh}(\frac{\|\Vec{v}\|_L}{\sqrt{K}})\Vec{x} + \sqrt{K}\text{sinh}(\frac{\| \Vec{v}\|_L}{\sqrt{K}})\frac{\Vec{v}}{\|\Vec{v}\|_L},\\
        &\texttt{log}_{\Vec{x}}(\Vec{y}) = \texttt{Dist}(\Vec{x}, \Vec{y}) \frac{\Vec{y} + \frac{1}{K}\langle \Vec{x}, \Vec{y}\rangle_L \Vec{x}}{\|\Vec{y} + \frac{1}{K}\langle \Vec{x}, \Vec{y}\rangle_L \Vec{x}\|_L},\\
        &\texttt{Dist}(\Vec{x}, \Vec{y}) = \sqrt{K}\text{arcosh}(-\langle \Vec{x}, \Vec{y}\rangle_L / K),
    \end{split}
\end{equation}
\noindent where $\Vec{v}\in\mathcal{T}_x\mathbb{H}^{d,K}$ and $\Vec{x}, \Vec{y}\in\mathbb{H}^{d,K}$. $\texttt{Dist}(\Vec{x}, \Vec{y})$ indicate the intrinsic distance between two points $\Vec{x}$ and $\Vec{y}$ in $\mathbb{H}^{d,K}$. We denote $\|\Vec{v}\|_L = \sqrt{\langle \Vec{v}, \Vec{v}\rangle_L}$ as the norm of $\Vec{v}\in\mathcal{T}_x\mathbb{H}^{d,K}$.

\subsubsection{Poincaré Ball Model}
The Poincaré Ball model with constant negative curvature ($c=-\frac{1}{K}, K>0$) is an open unit ball, \textit{i.e.}, $\mathbb{H}^{d,K}:=\{\Vec{x}\in\mathbb{R}^d: \|\Vec{x}\|<1\}$. For any point $\Vec{x}\in\mathbb{H}^{d,K}$, the exponential map $\texttt{exp}_{\Vec{x}}: \mathcal{T}_{\Vec{x}} \mathbb{
H}^{d,K}\rightarrow \mathbb{H}^{d,K}$ and the logarithmic map $\texttt{log}_{\Vec{x}}: \mathbb{H}^{d,K}\rightarrow  \mathcal{T}_{\Vec{x}}\mathbb{H}^{d,K}$ are defined, respectively, as:

\begin{eqnarray}
    \resizebox{0.91\hsize}{!}{%
    $\begin{split}
        &\texttt{exp}_{\Vec{x}}(\Vec{v}) = \Vec{x}\oplus \Big(\text{tanh}(\frac{\|\Vec{v}\|}{\sqrt{K}(1-\frac{1}{K}\|\Vec{x}\|^2)})\frac{\Vec{v}}{\frac{1}{\sqrt{K}}\|\Vec{v}\|}\Big),\\
        &\texttt{log}_{\Vec{x}}(\Vec{y}) = \sqrt{K}(1-\frac{1}{K}\|\Vec{x}\|^2)\text{arctanh}(\frac{1}{\sqrt{K}}\|-\Vec{x}\oplus \Vec{y}\|)\frac{-\Vec{x}\oplus \Vec{y}}{\|-\Vec{x}\oplus \Vec{y}\|},\\
        &\texttt{Dist}(\Vec{x}, \Vec{y}) = 2\sqrt{K} \text{arctanh}(\frac{1}{\sqrt{K}}\|-\Vec{x}\oplus \Vec{y}\|),
    \end{split}$
    }
\end{eqnarray}
\noindent where $\oplus$ is the Mobius addition for any $\Vec{x}, \Vec{y}\in\mathbb{H}^{d,K}$. Mobius addition is defined as,

\begin{eqnarray}
    \Vec{x}\oplus \Vec{y} = \frac{(1 + \frac{2}{K}\langle \Vec{x}, \Vec{y}\rangle + \frac{1}{K}\|\Vec{y}\|^2)\Vec{x} + (1 - \frac{1}{K}\|\Vec{x}\|^2)\Vec{y}}{1 + \frac{2}{K}\langle \Vec{x}, \Vec{y}\rangle + \frac{1}{K^2}\|\Vec{x}\|^2\|\Vec{y}\|^2},
\end{eqnarray}
\noindent with $\langle \cdot, \cdot\rangle$ being the inner product.

\subsection{Signed Neighbor Aggregation}\label{sec:agg}
In this section, we discuss how to aggregate node's information from its positive and negative neighborhood after obtaining the transformed features on the hyperbolic space. We first define the set of positive and negative neighbors of a user $i$ to be $\mathcal{N}^+(i)$ and $\mathcal{N}^-(i)$, respectively. For the first aggregation layer, we have 
\begin{equation}
\resizebox{0.91\hsize}{!}{%
$\begin{split}
     &\Vec{h}_i^{P(1),H} = \texttt{exp}_{\Vec{x}_i^H}\Big(\sigma\big(\Mat{W}^{(1)}_p[\sum_{j\in\mathcal{N}^+(i)}\Vec{w}^{+(1)}_{ij}\texttt{log}_{\Vec{o}}(\Vec{x}_j^H), \Vec{x}_i^H]\big)\Big),\\
     &\Vec{h}_i^{N(1),H} = \texttt{exp}_{\Vec{x}_i^H}\Big(\sigma\big(\Mat{W}^{(1)}_n[\sum_{j\in\mathcal{N}^-(i)}\Vec{w}^{-(1)}_{ij}\texttt{log}_{\Vec{o}}(\Vec{x}_j^H), \Vec{x}_i^H]\big)\Big),\\
     &\Vec{w}_{ij}^{+(1)} = \mathcal{F}_{j\in\mathcal{N}^+(i)}\Big(\sigma(\Vec{a}^{(1)T}_p[\Mat{W}^{(1)}_{pw} \texttt{log}_{\Vec{o}} \Vec{x}^H_i, \Mat{W}^{(1)}_{pw}  \texttt{log}_{\Vec{o}} \Vec{x}^H_j])\Big),\\
     &\Vec{w}_{ij}^{-(1)} = \mathcal{F}_{j\in\mathcal{N}^-(i)}\Big(\sigma(\Vec{a}^{(1)T}_n[\Mat{W}^{(1)}_{nw} \texttt{log}_{\Vec{o}} \Vec{x}^H_i, \Mat{W}^{(1)}_{nw}  \texttt{log}_{\Vec{o}} \Vec{x}^H_j])\Big),
\end{split}$
}
\end{equation}
where the function $\mathcal{F}_{j\in\mathcal{N}(i)}(\Vec{x}) = (2e^{\Vec{x}_i}/\sum_{j} e^{\Vec{x}_j})-1$ regularizes the attention weights $w_{ij}^{+(1)}, w_{ij}^{-(1)}$ in a range of $(-1, 1)$. $\Mat{W}^{(1)}_p, \Mat{W}^{(1)}_n\in\mathbb{R}^{d\times 2d}$, $\Vec{a}^{(1)}_p, \Vec{a}^{(1)}_n\in\mathbb{R}^{2d\times d}$, and $\Mat{W}^{(1)}_{pw}, \Mat{W}^{(1)}_{nw}\in\mathbb{R}^{d\times d}$ are the weight matrices. $\sigma(\cdot)$ is the LeakyReLU activation, with $[\cdot,\cdot]$ being the concatenation operation. The hidden representation $\Vec{h}_i^{P(1),H}\in\mathbb{H}^{d,K}$ and $\Vec{h}_i^{N(1),H}\in\mathbb{H}^{d,K}$ collect the positive and negative supports from the adjacent positive \textcolor{black}{and negative} neighbors, respectively, which allow to embed node features close to positive neighbors and distant to negative neighbors. However, for multi-hop neighborhood, the relationships become complex to determine. According to Figure \ref{fig:theory}, the aggregation rule for $l>1$ is defined as,

\begin{equation}
\resizebox{0.91\hsize}{!}{%
    $\begin{split}
     &\hat{\Vec{h}}_i^{P(l),H} = [\sum_{j\in\mathcal{N}^+(i)}\Vec{w}^{+(l)}_{ij}\texttt{log}_{\Vec{o}}(\Vec{h}_j^{P(l-1),H}), \sum_{k\in\mathcal{N}^-(i)}\Vec{w}^{-(l)}_{ik}\texttt{log}_{\Vec{o}}(\Vec{h}_k^{N(l-1),H})],\\
     &\hat{\Vec{h}}_i^{N(l),H} = [\sum_{j\in\mathcal{N}^-(i)}\Vec{w}^{-(l)}_{ij}\texttt{log}_{\Vec{o}}(\Vec{h}_j^{N(l-1),H}), \sum_{k\in\mathcal{N}^+(i)}\Vec{w}^{+(l)}_{ik}\texttt{log}_{\Vec{o}}(\Vec{h}_k^{P(l-1),H})],\\
     &\Vec{h}_i^{P(l),H} = \texttt{exp}_{\Vec{h}_i^{P(l-1),H}}\Big(\sigma\big(\Mat{W}^{(l)}_p[\hat{h}_i^{P(l),H}, \Vec{h}_i^{P(l-1),H}]\big)\Big),\\
     &\Vec{h}_i^{N(l),H} = \texttt{exp}_{\Vec{h}_i^{N(l-1),H}}\Big(\sigma\big(\Mat{W}^{(l)}_n[\hat{h}_i^{N(l),H}, \Vec{h}_i^{N(l-1),H}]\big)\Big).
    \end{split}$
    }
\end{equation}
The node latent feature will be updated iteratively with the weighted aggregation of neighbors' features. The weight factors for positive and negative neighbors' features are calculated as,
\begin{equation}
\resizebox{0.91\hsize}{!}{%
    $\begin{split}
 &w_{ij}^{+(l)} = \mathcal{F}_{j\in\mathcal{N}^+(i)}\Big(\sigma(\Vec{a}^{(l)T}_p[\Mat{W}^{(l)}_{pw} \texttt{log}_{\Vec{o}} \Vec{h}_i^{P(l-1),H}, \Mat{W}_{pw}^{(l)}  \texttt{log}_{\Vec{o}} \Vec{h}_j^{P(l-1),H}])\Big),\\
     &w_{ij}^{-(l)} = \mathcal{F}_{j\in\mathcal{N}^-(i)}\Big(\sigma(\Vec{a}^{(l)T}_n[\Mat{W}^{(l)}_{nw} \texttt{log}_{\Vec{o}} \Vec{h}_i^{N(l-1),H}, \Mat{W}_{nw}^{(l)}  \texttt{log}_{\Vec{o}} \Vec{h}_j^{N(l-1),H}])\Big),\\
     &w_{ik}^{+(l)} = \mathcal{F}_{k\in\mathcal{N}^+(i)}\Big(\sigma(\Vec{a}^{(l)T}_p[\Mat{W}^{(l)}_{nw} \texttt{log}_{\Vec{o}}\Vec{h}_i^{N(l-1),H}, \Mat{W}_{pw}^{(l)}  \texttt{log}_{\Vec{o}} \Vec{h}_k^{P(l-1),H}])\Big),\\
     &w_{ik}^{-(l)} = \mathcal{F}_{k\in\mathcal{N}^-(i)}\Big(\sigma(\Vec{a}^{(l)T}_n[\Mat{W}^{(l)}_{pw} \texttt{log}_{\Vec{o}} \Vec{h}_i^{P(l-1),H}, \Mat{W}_{nw}^{(l)}  \texttt{log}_{\Vec{o}} \Vec{h}_k^{N(l-1),H}])\Big),
    \end{split}$
    }
\end{equation}
% \end{equation}
\noindent with $\mathcal{F}_{j\in\mathcal{N}(i)}(\Vec{x}) = (2e^{\Vec{x}_i}/\sum_{j} e^{\Vec{x}_j})-1$. Similarly, $\Mat{W}^{(l)}_p, \Mat{W}^{(l)}_n\in\mathbb{R}^{d\times 2d}$, $\Vec{a}^{(l)}_p, \Vec{a}^{(l)}_n\in\mathbb{R}^{2d\times d}$ and $\Mat{W}^{(l)}_{pw}, \Mat{W}^{(l)}_{nw}\in\mathbb{R}^{d\times d}$ are learnable weight matrices. The final representation $\Vec{z}_i^H\in\mathbb{H}^{d,K}$ for each node $i$ at the $L$-th layer can be obtained by,
\begin{equation}
    \Vec{z}_i^H = [\Vec{h}_i^{P(L),H}, \Vec{h}_i^{N(L),H}].
\end{equation}

\subsection{Mutual Information Maximization}\label{sec:mim}
For guiding the path of aggregation between balance theory and status theory, a Shannon entropy-based measure, \textit{i.e.}, Mutual information (MI) is leveraged to measure the correlations between the adjacency matrix $\Mat{A}$ and the concatenated representations of node-pairs $\Mat{Z}$, with $\Vec{z}_{ij} = [\Vec{z}_i^H, \Vec{z}_j^H]\in\mathbb{H}^{2d,K}$. \textcolor{black}{By maximizing the mutual information, it is anticipated that the models can identify a subset of node neighbors that are most influential for the edge prediction during the message passing and determine either push the neighbors' features closer or further in the node embedding space. To measure the mutual dependence between the two variables $\Vec{Z}$ and $\Mat{A}$, we first introduce some notations for calculation. Let ($\Vec{Z}$, $\Mat{A}$) be a pair with values over the space $\mathcal{Z}\times\mathcal{A}$. The joint distribution is denoted by $\mathbb{P}_{\Mat{Z}\Mat{A}}$ and the marginal distributions are $\mathbb{P}_{\Vec{Z}}$ and $\mathbb{P}_{\Mat{A}}$, respectively.} Based on the principle of \cite{infomax,mine}, the mutual information $I(\Mat{Z}, \Mat{A})$ is equivalent to the Kullback-Leibler (KL) divergence between the joint and product of the marginals, $\mathbb{P}_{\Vec{Z}\Mat{A}}$ and $\mathbb{P}_{\Vec{Z}}\otimes\mathbb{P}_{\Mat{A}}$:
{\color{black}
\begin{equation}\label{eq:mi}
    \begin{split}
    I(\Mat{Z}, \Mat{A}) &= \infdiv{\mathbb{P}_{\Vec{Z}\Mat{A}}}{\mathbb{P}_{\Vec{Z}}\otimes\mathbb{P}_{\Mat{A}}},\\
    &=\mathbb{E}_{\mathbb{P}_{\Vec{Z}\Mat{A}}}[\log \frac{d\mathbb{P}_{\Vec{Z}\Mat{A}}}{d\mathbb{P}_{\Vec{Z}\otimes\Mat{A}}}].\\
    \end{split}
\end{equation}
\noindent Here we consider $\mathbb{P}_{\Vec{Z}\Mat{A}}$ amd $\mathbb{P}_{\Vec{Z}\otimes\Mat{A}}$ as being distributions on a compact domain $\Omega\in\mathbb{R}^{2d}$. Equation \eqref{eq:mi} can be interpreted as, the larger the divergence between the joint and the product of the marginals, the stronger the dependence between $\Mat{Z}$ and $\Mat{A}$.} Maximizing mutual information forces the node embeddings to aggregate features from their positive and negative neighbors within $L$ hops. {\color{black} However, both the joint distribution and marginal distributions are intractable for optimization, which motivates various approaches to estimate the tractable lower bound of mutual information. One of the most commonly used is the Donsker-Varadhan representation~\cite{DV} of the KL-divergence, with which the lower-bound to the mutual information $\hat{I}_\theta(\Mat{Z}, \Mat{A})$ can be derived as,
\begin{equation}\label{eq:mim}
\begin{split}
   \hat{I}_\theta(\Mat{Z}, \Mat{A}):= \underset{T:\Omega\rightarrow\mathbb{R}}{\text{sup}}\mathbb{E}_{\mathbb{P}_{\Vec{Z}\Mat{A}}}[T_{\Vec{\theta}}] - \log(\mathbb{E}_{\mathbb{P}_{\Vec{Z}}\otimes\mathbb{P}_{\Mat{A}}}[e^{T_\theta}]),
\end{split}
\end{equation}
\noindent where $T_{\Vec{\theta}}: \Mat{Z}\times \Mat{A}\rightarrow \mathbb{R}$ is a discriminator function parameterized by $\Vec{\theta}$, which takes the $\Vec{Z}$ and $\Mat{A}$ as inputs and turns a value vector. More details of $T_{\Vec{\theta}}$ can be found in Section \ref{sec:implement}. The expectations in Equation \eqref{eq:mim} are finite and can be estimated using \textit{i.i.d} samples from the joint distribution and shuffled samples from the respective marginal distributions, respectively. }The objective function in Equation~\eqref{eq:obj} can be maximized by gradient ascent, as:
\begin{equation}\label{eq:obj}
    \mathcal{L}_{mim} = -\hat{I}_\theta(\Mat{Z}, \Mat{A}).
\end{equation}

\subsection{Edge Classifier and Training Objectives}
For the task of signed link prediction, we use the Fermi-Dirac decoder to generate the predictions, which are supervised by a binary cross entropy loss:
\begin{equation}
\begin{split}
     &p_{ij} = [e^{(\texttt{Dist}(\Vec{z}_i^H, \Vec{z}_j^H)^2 - r) / t}\color{black} + 1\color{black}]^{-1},\\
     &\mathcal{L}_{cls} = \sum_{i,j\in\mathcal{V}}\Mat{A}_{ij} \text{log}(p_{ij}) + (1-\Mat{A}_{ij})\text{log}(1-p_{ij}),
\end{split}
\end{equation}
with $r$ and $t$ being the hyperparameters. To further constrain the node embeddings, we design a positive ranking loss and a negative ranking loss, respectively. For each positive or negative pair of users $(i, j)$, we randomly sample a neutral user $k$ which has no link to the anchor $i$. The following objectives enable the positively linked users closer (and negatively linked users farther) in the embedded space than the no-link pairs $(i, k)$:  
\begin{equation}
\begin{split}
    \mathcal{L}_{pos} = \sum_{(i,j)\in\mathcal{E}^+, k} max(0, \texttt{Dist}(\Vec{z}_i^H, \Vec{z}_j^H) - \texttt{Dist}(\Vec{z}_i^H, \Vec{z}_k^H)),\\
    \mathcal{L}_{neg} = \sum_{(i,j)\in\mathcal{E}^-, k} max(0, \texttt{Dist}(\Vec{z}_i^H, \Vec{z}_k^H) - \texttt{Dist}(\Vec{z}_i^H, \Vec{z}_j^H)).
\end{split}
\end{equation}
Lastly, the model is jointly trained by edge classification loss, mutual information loss, and two embedding losses:
\begin{equation}
    \mathcal{L} = \mathcal{L}_{cls} + \alpha \mathcal{L}_{pos} + \beta \mathcal{L}_{neg} + \gamma \mathcal{L}_{mim},
\end{equation}
where $\alpha$, $\beta$, and $\gamma$ denote the loss coefficients, respectively.
\section{Experimental Settings}\label{sec:setting}
\subsection{Datasets and Evaluation Metrics}
In this section, we conduct extensive experiments on four real-world signed social network datasets, \textit{i.e.}, Bitcoin-Alpha\footnote{http://www.btcalpha.com/}, Bitcoin-OTC\footnote{http://www.bitcoin-otc.com/}, Slashdot\footnote{http://slashdot.org/}, Epinions\footnote{http://www.epinions.com/}. The general statistics of the four network datasets are summarized in Table \ref{tab:datasets}
.
\begin{itemize}
    \item \textbf{Bitcoin-Alpha} and \textbf{Bitcoin-OTC}~\cite{bitcoin,bitcoin1} are two who-trusts-whom networks of Bitcoin trading. Members of Bitcoin Alpha and OTC rate other members as trust or distrust to prevent transactions from risky users.
    \item \textbf{Slashdot}~\cite{slashdot} is collected from a technology-related news website known for its specific user community. The website features user-submitted and editor-evaluated technology oriented news. It allows users to tag each other as friends (positive links) or foes (negative links). 
    \item \textbf{Epinions}~\cite{slashdot} is a trust network for the consumer review site. All the trust relationships interact and form the Web of Trust, which is then combined with review ratings to determine which reviews are shown to the user.
\end{itemize}
\noindent In order to predict the link relationship of the unconnected node pairs, we randomly select $20\%$ of the links in the social networks to form a test set, with the remaining links as the training set. We utilize the standard metrics \textit{i.e.}, area under curve (\textbf{AUC}), \textbf{F1} score, \textbf{macro-averaged F1} score, and \textbf{micro-averaged F1} score to evaluate the prediction performance.

\vspace{-2ex}
\subsection{Baselines Methods}
We compare our approach with the following signed network embedding and link prediction methods:
\begin{itemize}
    \item \textbf{TSVD}~\cite{tsvd}: performs linear dimensionality reduction by means of truncated singular value decomposition (SVD).
    \item \textbf{SSE}~\cite{SSE}: reformulates the Rayleigh quotient as an objective for embedding learning.
    \item \textbf{SiNE}~\cite{SiNE}: optimizes an objective function guided by social theory in signed networks to generate the node embeddings in a deep learning framework.
    \item \textbf{SIDE}~\cite{SIDE}: provides a linearly scalable method to obtain the low-dimensional vectors with random walks.
    \item \textbf{SIGNet}~\cite{SIGNet}: builds upon word2vec embedding approaches and adds a sampling strategy to maintain balance in high-order neighborhoods.
    \item \textbf{SGCN}~\cite{SGCN}: generalizes GNNs to a signed network for the first time. It designs a new aggregation strategy for undirected signed network. 
    \item \textbf{SiGAT}~\cite{SiGAT}: incorporates graph motifs into GAT to capture the balance theory and status theory jointly.
    \item \textbf{SNEA}~\cite{SNEA}: proposes a graph attention layer to estimate the importance coefficients for the node pairs.
\end{itemize}

\subsection{Implementation Details}\label{sec:implement}
Our source code is based on PyTorch \cite{pytorch}, which is available in an anonymous repository\footnote{\textcolor{black}{https://github.com/Luoyadan/SIHG}} for reference. All experiments are conducted on two servers with two GeForce GTX 2080 Ti GPUs. Similar to previous works in this area~\cite{SGCN}, random seed is set to 42. For fair comparisons, the feature dimensions of node embedding $d$ are fixed to 64. The total number of training epochs $M$ is 800 for Bitcoin-Alpha and Bitcoin-OTC datasets, 900 for Slashdot and Epinions datasets. The Adam optimizer is applied with a weight decay of $1\times10^{-5}$. The learning rate $\mu$ is initiated to be $1\times10^{-2}$ for Bitcoin-Alpha and Bitcoin-OTC datasets, and $5\times10^{-3}$ for Slashdot and Epinions datasets. The learning rate is adapted by a cosine annealing schedule. The node embedding is initialized with TruncatedSVD~\cite{tsvd}, with maximum 30 iterations. The optimal loss coefficients $\alpha$, $\gamma$ are searched with optuna~\cite{optuna} framework, and $\beta$ is empirically set to 0.83 for all tasks. The hyperparameter $r$ and $t$ for edge classifier are fixed to 2 and 1, respectively. Without loss of generality, we set the curvature $K=1$ (\textit{i.e.} $c=-1$), which can be further tuned for $\mathcal{G}_p$ and $\mathcal{G}_n$. \textcolor{black}{The $T_{\Vec{\theta}}$ network firstly maps each node-pair representation / edge sign from $128$-D / $1$-D to $128$-D with two fully connected layers. It then projects the addition of the two 128-D vectors to 1-D scores with a LeakyReLU and a fully connected layer.}

\begin{table}[t]
\centering 
\caption{The general statistics of the four datasets used in our experiments.}
\scalebox{0.9}{% 
\begin{tabular}{l  c c c c c}
\toprule % Top horizontal line
Datasets &\# Nodes &\# Links &\% Positive Links &\% Negative Links \\ % Column names row
\midrule % In-table horizontal line
\midrule
Bitcoin-Alpha &3,783 &14,145 &89.99 &10.01\\ % Content row 1
Bitcoin-OTC &5,881 &21,522 &85.45 &14.55\\
Slashdot &82,140 &549,202 &77.40 &22.60\\
Epinions &131,827  &841,372 &85.30 &14.70\\
\bottomrule % Bottom horizontal line
\end{tabular}
}
\label{tab:datasets}
\end{table}

\vspace{-2ex}
\section{Experimental Results and Analysis}
Following the settings in Section \ref{sec:setting}, we conduct experiments to evaluate the performance of the proposed SIHG regarding both the signed link prediction effectiveness and interpretation quality. In particular, we aim to answer the following research questions (RQs) via experiments:\\
\textbf{RQ1:} How effectively can SIHG perform signed link prediction compared with state-of-the-art baselines? \\
\textbf{RQ2:} What is the contribution of each key component of the proposed model structure?\\
\textbf{RQ3:} How the hyperparameters affect the performance of SIHG in terms of prediction effectiveness?\\
\textbf{RQ4:} How is the quality of the learned node representations?\\
\textbf{RQ5:} How to interpret the underlying social theories with the learned signed attention maps?

\vspace{-2ex}
\subsection{Signed Link Prediction Effectiveness (RQ1)}
In Table \ref{tab:auc}, we report the signed link prediction results across the four benchmark datasets in terms of \textbf{AUC} and \textbf{F1} scores. The baseline results refer to \cite{SNEA,SGCN}. The graph depth $L$ is fixed to 3 and the Hyperboloid model is incorperated in \textbf{SIHG}. In order to fully investigate the graph-based baselines that are highly related to our work, we re-implement \textbf{SGCN}, \textbf{SiGAT}, and \textbf{SNEA}, and additionally report the \textbf{macro-F1} and \textbf{micro-F1} scores in Table \ref{tab:f1}. It is observed that the proposed SIHG framework is superior to all the compared methods in most cases. Among the four commonly-used evaluation metrics, the AUC scores of SIHG are boosted by the largest margin ($13.03\%$ on Slashdot) over the best performing baseline, especially on the large-scale datasets, while the macro-F1 scores are slightly weaker. We infer this result is due to using macro-average, which computes the score independently for each class (\textit{i.e.} $-1, 1$) and treats all classes equally. Therefore, using macro-F1 cannot fairly testify the prediction quality on extremely biased datasets such as signed social networks.

\vspace{-2ex}

\begin{table}[t]
\centering 
\caption{AUC and F1 scores of predicting signed edges among four datasets.}
\resizebox{1\linewidth}{!}{% 
\begin{tabular}{l c c c c c c c c c}
\toprule % Top horizontal line
\multirow{2}{*}{Method} &\multicolumn{2}{c}{\textbf{Bitcoin-Alpha}} &\multicolumn{2}{c}{\textbf{Bitcoin-OTC}} &\multicolumn{2}{c}{\textbf{Slashdot}} &\multicolumn{2}{c}{\textbf{Epinions}}\\ 
\cmidrule(l){2-3}\cmidrule(l){4-5} \cmidrule(l){6-7} \cmidrule(l){8-9}
&AUC &F1 &AUC &F1 &AUC &F1 &AUC &F1\\
\midrule % In-table horizontal line
\midrule
TSVD~\cite{tsvd} &0.740 &0.863 &0.761 &0.870 &0.740 &0.804 &0.766 &0.843\\
SSE~\cite{SSE} &0.764 &0.898 &0.803 &0.923 &0.769 &0.820 &0.822 &0.901\\
SiNE~\cite{SiNE} &0.781 &0.895 &0.782 &0.876 &0.785 &0.850 &0.831 &0.902\\
SIDE~\cite{SIDE} &0.642 &0.753 &0.632 &0.728 &0.554 &0.624 &0.617  &0.725\\
SGCN~\cite{SGCN} &0.801 &0.915 &0.804 &0.908 &0.786 &0.859 &0.849 &0.920\\
SiGAT~\cite{SiGAT} &0.775 &0.894 &0.796 &0.903 &0.789 &0.857 &0.853 &0.917\\
% SNEA-1 &0.766 &0.784 &0.726 &0.822\\
SNEA~\cite{SNEA} &0.816 &0.927 &0.818 &0.924 &0.799 &0.868 &0.861 &0.933\\
\midrule
\midrule
SIHG &\textbf{0.898} &\textbf{0.961} &\textbf{0.915} &\textbf{0.953} &\textbf{0.895}  &\textbf{0.919}  &\textbf{0.926} &\textbf{0.957}\\ 
\bottomrule % Bottom horizontal line
\end{tabular}
}
\label{tab:auc}
\end{table}

\begin{table} 
\centering 
\caption{The signed link prediction results of the proposed method and compared baselines. $^*$ indicates the results of re-implrementation.}
\resizebox{1\linewidth}{!}{% 
\begin{tabular}{l  l c c c c c}
\toprule % Top horizontal line
Dataset &Method &AUC &F1 &macro-F1 &micro-F1\\
\midrule
\multirow{4}{*}{Bitcoin-Alpha} &SGCN$^*$~\cite{SGCN} &0.8147 &0.8996 &0.6836 &0.8310\\ % 
&SiGAT$^*$~\cite{SiGAT} &0.8393 &0.9519 &0.6721 &0.9109 \\
&SNEA$^*$~\cite{SNEA} &0.8293 &0.9297 &\textbf{0.7430} &0.8786\\
&SIHG &\textbf{0.8981} &\textbf{0.9614} &0.7115 &\textbf{0.9279}\\
% &+0.070 &
\midrule
\multirow{4}{*}{Bitcoin-OTC} &SGCN$^*$~\cite{SGCN} &0.8087 &0.9152 &0.7617 &0.8605\\ % 
&SiGAT$^*$~\cite{SiGAT} &0.8797 &0.9423 &0.7578 &0.8983\\
&SNEA$^*$~\cite{SNEA} &0.8131 &0.9174 &0.7705 &0.8646 \\
&SIHG &\textbf{0.9154} &\textbf{0.9528}  &\textbf{0.7949} &\textbf{0.9165}\\
% &+0.041
\midrule
\multirow{4}{*}{Slashdot} &SGCN$^*$~\cite{SGCN} &0.7827 &0.8688 &0.7512 &0.8068\\ % 
&SNEA$^*$~\cite{SNEA} &0.7918 &0.8627 &0.7634 &0.8051\\
&\textcolor{black}{SIGNet~\cite{SIGNet}} &- &- &- &\textcolor{black}{0.8320}\\
&SIHG &\textbf{0.8950} &\textbf{0.9189} &\textbf{0.7934} &\textbf{0.8696} \\
% &+0.1303
\midrule
\multirow{4}{*}{Epinions} &\textcolor{black}{SGCN$^*$~\cite{SGCN}} &\textcolor{black}{0.8343} &\textcolor{black}{0.8001} &\textcolor{black}{0.6277} &\textcolor{black}{0.7075}\\ % 
&SNEA$^*$~\cite{SNEA} &0.8542 &0.9304 &0.8167 &0.8873 \\
&\textcolor{black}{SIGNet~\cite{SIGNet}} &- &- &- &\textcolor{black}{0.9200}\\
&SIHG &\textbf{0.9262} &\textbf{0.9571} &\textbf{0.8261} &\textbf{0.9247} \\
% &0.084
\bottomrule
\end{tabular}
}
\label{tab:f1}
\end{table}

\begin{table} 
\centering 
\captionsetup{labelfont={color=black},font={color=black}}
\caption{The study of node embedding size. $^*$ indicates the results of re-implrementation.}
\resizebox{1\linewidth}{!}{% 
\begin{tabular}{>{\color{black}}l >{\color{black}}l >{\color{black}}c >{\color{black}}c >{\color{black}}c >{\color{black}}c >{\color{black}}c}
\toprule % Top horizontal line
Dataset &Method &AUC &F1 &macro-F1 &micro-F1\\
\midrule
\multirow{6}{*}{Bitcoin-Alpha} 
&SGCN$^*$-32 &0.8111 &0.7778 &0.5440 &0.6638\\
&SGCN$^*$-64 &0.8147 &0.8996 &0.6836 &0.8310\\  
&SGCN$^*$-128 &0.8310 &0.8972 &0.6621 &0.8257\\
&SIHG-32 &0.8867 &0.9587 &0.6714 &0.9226\\
&SIHG-64 &0.8981 &\textbf{0.9614} &\textbf{0.7115} &\textbf{0.9279}\\
&SIHG-128 &\textbf{0.9015} &0.9599 &0.6986 &0.9251\\
% &+0.1303
\midrule
\multirow{6}{*}{Bitcoin-OTC} 
&SGCN$^*$-32 &0.7942 &0.7831 &0.5984 &0.6833\\  
&SGCN$^*$-64 &0.8087 &0.9152 &0.7617 &0.8605\\
&SGCN$^*$-128 &0.8751 &0.9278 &0.7792 &0.8792\\
&SIHG-32  &0.9041 &0.9496 &0.7744 &0.9104\\
&SIHG-64  &0.9154 &\textbf{0.9528} &0.7949 &\textbf{0.9165}\\
&SIHG-128 &\textbf{0.9200} &0.9513 &\textbf{0.7959} &0.9142\\
% &0.084
\bottomrule
\end{tabular}
}
\label{tab:embed_size}
\end{table}
\subsection{Ablation Study (RQ2)}
To investigate the validity of the derived MI objective (\textbf{MIM}), the signed attention mechanism (\textbf{Signed Attention}) and the incorporated Hyperboloid models (\textbf{Hyperbolic}), we compare the seven variants of SIHG model on both \textbf{Bitcoin-Alpha} and \textbf{Bitcoin-OTC} datasets, and summarize the comparison results in Table \ref{tab:abla}. The graph depth $L$ is fixed to 3 and loss coefficients $\alpha$ and $\gamma$ are empirically fixed. The first row corresponds to the simplest baseline, which resembles SGCN yet it has separate transformations for positive and negative neighbors' features. By comparing the rest of variants with one or two components removed, we observed degradation in the respective performances of signed link prediction. It is noteworthy that removing the MI objective will lead to a significant drop in prediction accuracy, which verifies the importance of discovering correlations between the edge polarity and the latent node embedding pairs.

\begin{table}
  \caption{Ablation study of the proposed SIHG on Bitcoin-Alpha and Bitcoin-OTC datasets.}
  \label{tab:abla}
  \resizebox{1\linewidth}{!}{% 
  \begin{tabular}{ccc|cccc}
    \toprule
    & & & \multicolumn{2}{c}{\textbf{Bitcoin-Alpha}} & \multicolumn{2}{c}{\textbf{Bitcoin-OTC}}\\
    \cmidrule(l){4-5} \cmidrule(l){6-7}
    MIM  &Signed Attention &Hyperbolic  &AUC &F1      &AUC &F1\\
    \hline
    -                             &-         &-         &0.8149 &0.9020 &0.8268 &0.9297\\
    \hline

    $\surd$                       &-         &-         &0.8867 &0.9602 &0.9071 &0.9493\\
    -                             &$\surd$   &-         &0.8532 &0.9571 &0.8925 &0.9456\\
    -                             &-         &$\surd$   &0.8756 &0.9580 &0.9004 &0.9517\\
    -                             &$\surd$   &$\surd$   &0.8767 &0.9588 &0.9034 &0.9510\\
    $\surd$                       &$\surd$   &-         &0.8880 &0.9597 &0.9115 &0.9507 \\
    $\surd$                       &-         &$\surd$   &0.8817 &0.9575 &0.9133 &0.9527\\
    \hline
    $\surd$                       &$\surd$   &$\surd$   &\textbf{0.8981} &\textbf{0.9614} &\textbf{0.9154} &\textbf{0.9528}\\
  \bottomrule
\end{tabular}
}
\end{table}
 
\begin{table}[t] 
\centering % Centres the table on the page, comment out to left-justify
\caption{The link prediction performances of the proposed SIHG w.r.t. various hyerbolic models on 3 datasets.}
\resizebox{1\linewidth}{!}{% 
\begin{tabular}{l l c c c c } 
\toprule % Top horizontal line
Dataset &Method &AUC &F1 &macro-F1 &micro-F1 \\ 
\midrule % In-table horizontal line
\multirow{3}{*}{Bitcoin-Alpha} &Euclidean &0.8880 &0.9597 &0.6934 &0.9247 \\ 
&Poincaré &0.8860 &0.9600 &0.6894 &0.9251 \\
&Hyperboloid &\textbf{0.8981} &\textbf{0.9614} &\textbf{0.7115} &\textbf{0.9279}\\
\midrule
\midrule
\multirow{3}{*}{Bitcoin-OTC} &Euclidean &0.9115 &0.9507 &0.7822 &0.9125\\
&Poincaré &0.9071 &0.9510 &0.7826 &0.9130\\
&Hyperboloid &\textbf{0.9154} &\textbf{0.9528}  &\textbf{0.7949} &\textbf{0.9165}\\
\bottomrule % Bottom horizontal line
\vspace{-2ex}
\end{tabular}
}
\label{tab:model} 
\end{table}

\subsection{Parameter Sensitivity (RQ3)}
We evaluate the sensitivity of the proposed SIHG method \textit{w.r.t} the choices of hyperbolic models, loss coefficients $\alpha$ and $\gamma$, and the depth $L$ of graph networks on the Bitcoin-Alpha and Bitcoin-OTC datasets.

\noindent\textit{\textbf{Effect of Hyperbolic Model.}} First, we assessed the impact of the base model of the SIHG, \textit{i.e.}, \textbf{Euclidean}, \textbf{Poincaré}, and \textbf{Hyperboloid} models w.r.t four evaluation metrics. Notably, for Euclidean model, we do not apply the $\texttt{log}$ and $\texttt{exp}$ maps. The results, reported in Table \ref{tab:model}, demonstrate that the proposed SIHG framework with the hyperboloid model achieve a relatively higher performance compared to the Euclidean alternative, improving AUC scores from $88.8\%$ to $89.8\%$. The performance improvement also illustrates the numerical stability of the Hyperboloid over the Poincaré.
\smallskip

\begin{figure}[t]
    \centering
    \subfloat[][Bitcoin-Alpha]{\includegraphics[width=0.5\linewidth]{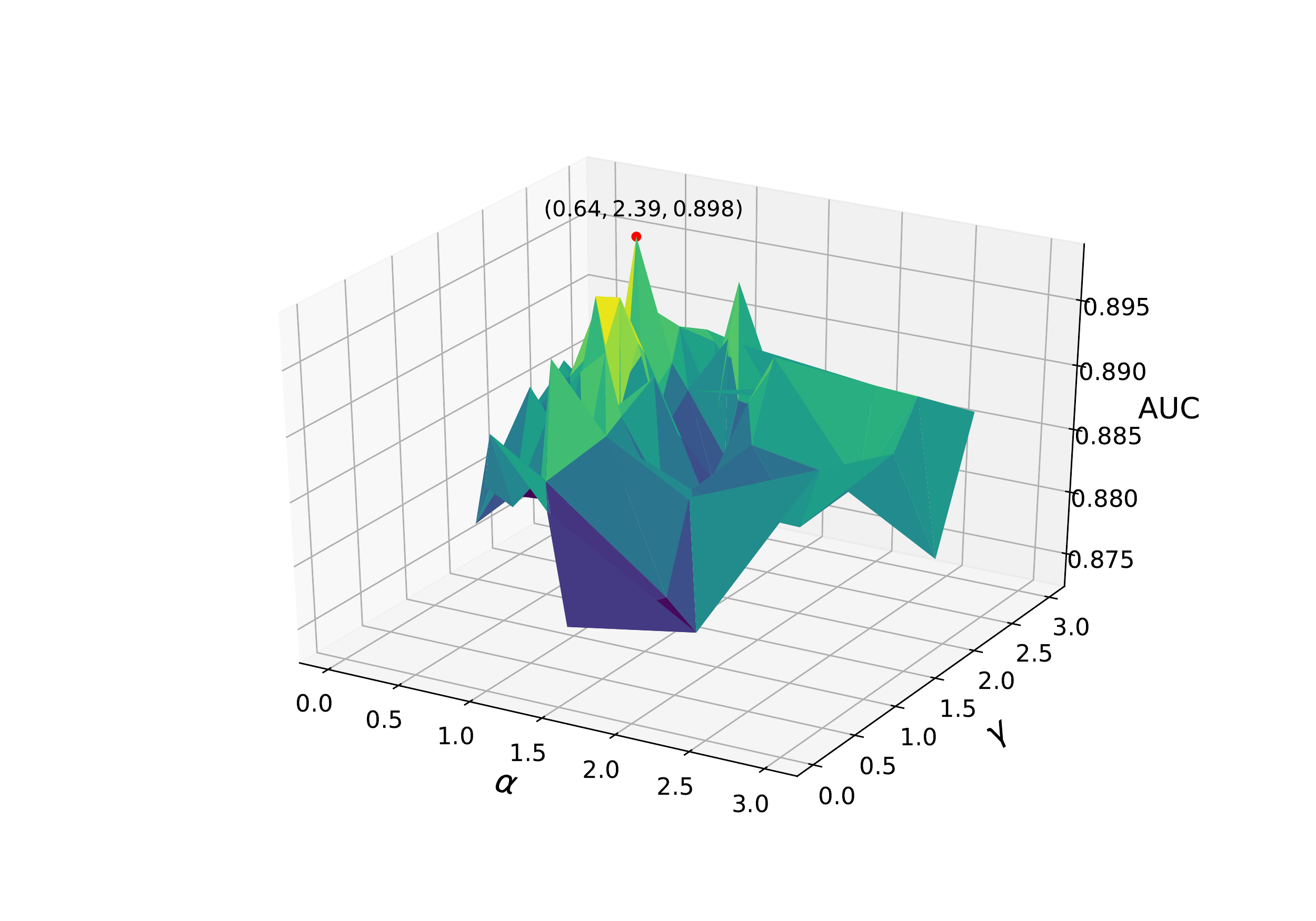}
    \includegraphics[width=0.5\linewidth]{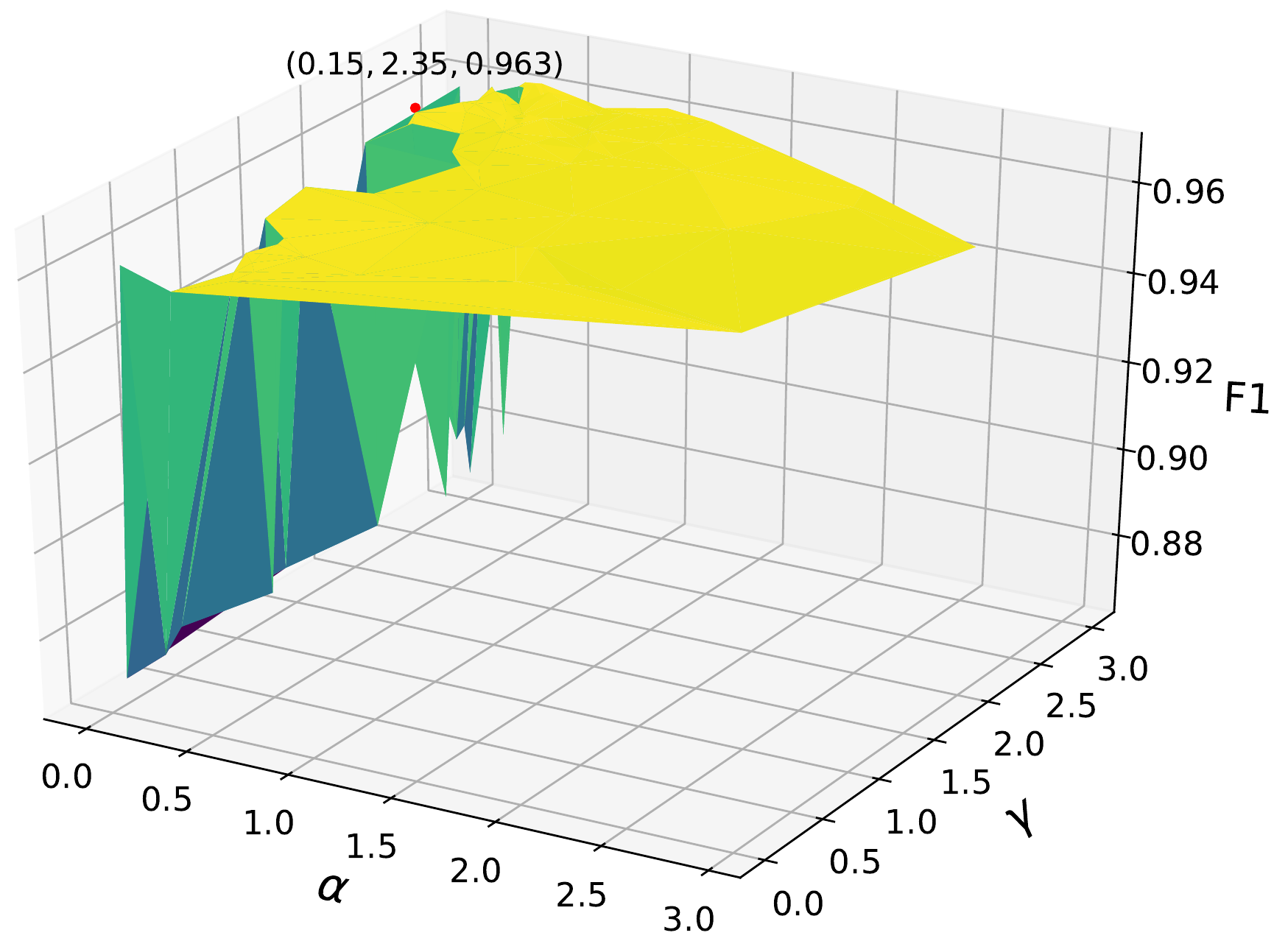}}\\
    \subfloat[][Bitcoin-OTC]{\includegraphics[width=0.5\linewidth]{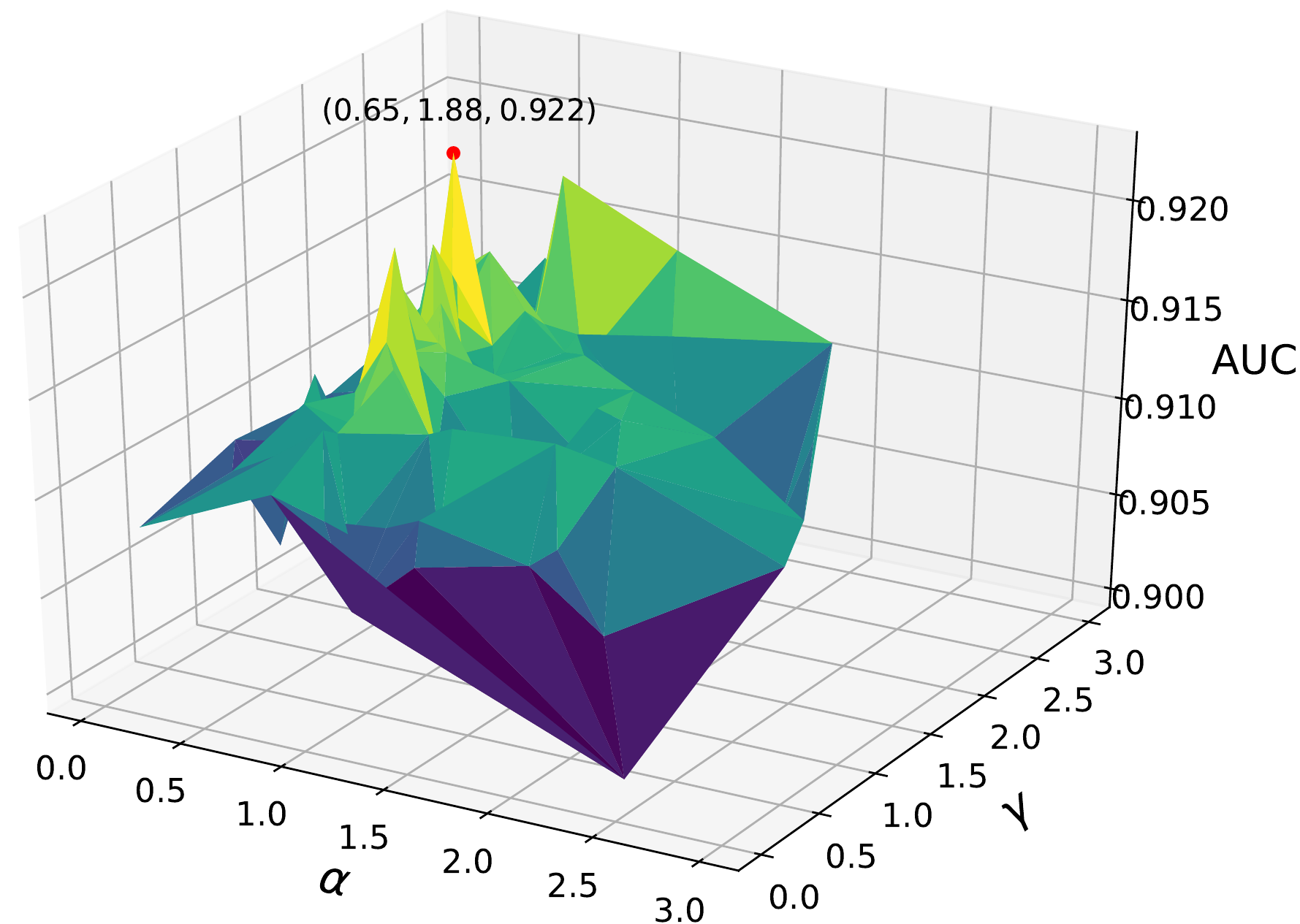}
    \includegraphics[width=0.5\linewidth]{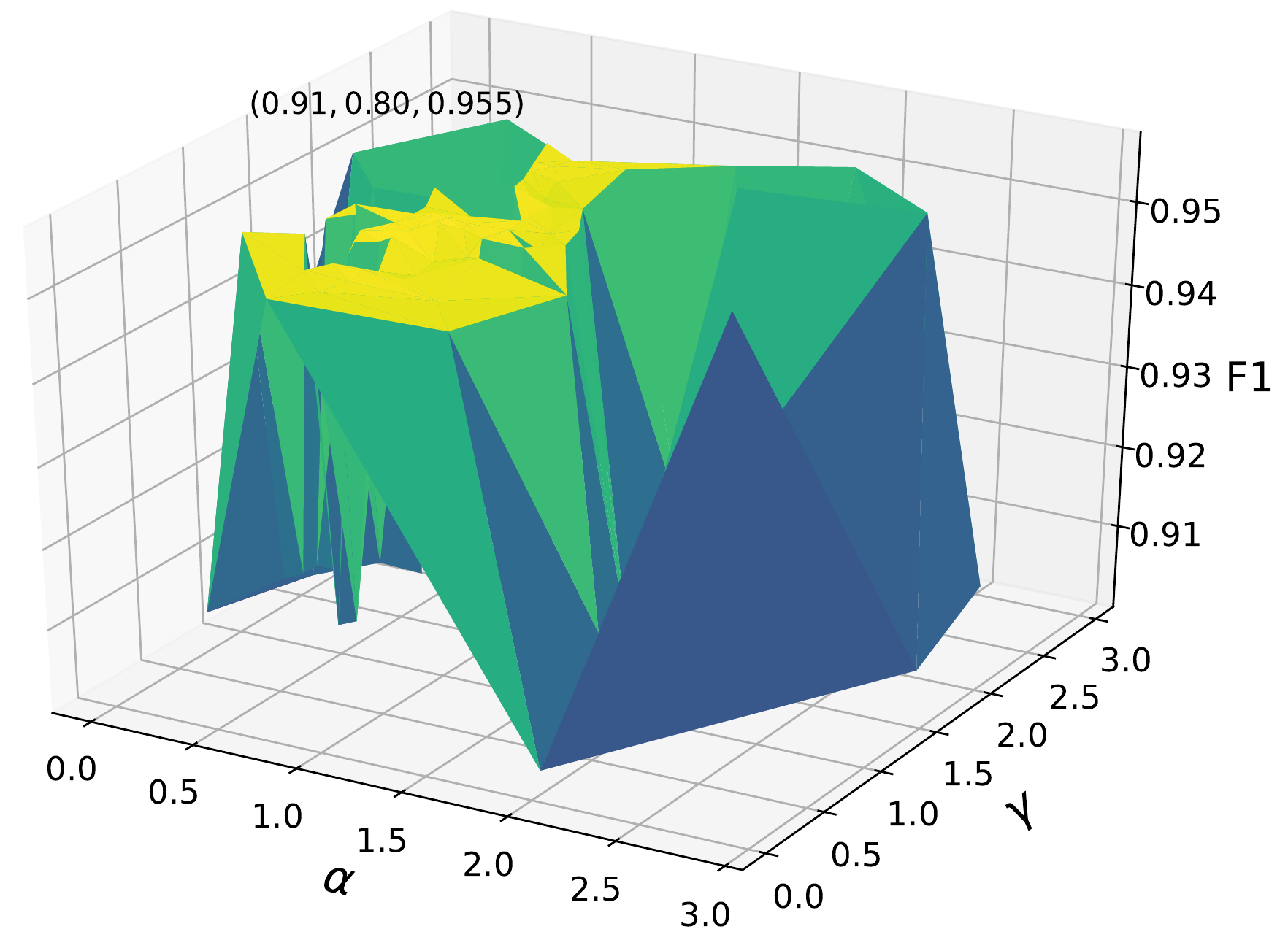}}
    \caption{The impact of loss coefficients $\alpha$ and $\gamma$ on signed link prediction. The best trial is marked with a red point.}
    \label{fig:params}
\end{figure}

\begin{figure}[t]\vspace{-3ex}
    \centering
    \subfloat[][Bitcoin-Alpha]{\includegraphics[width=0.8\linewidth]{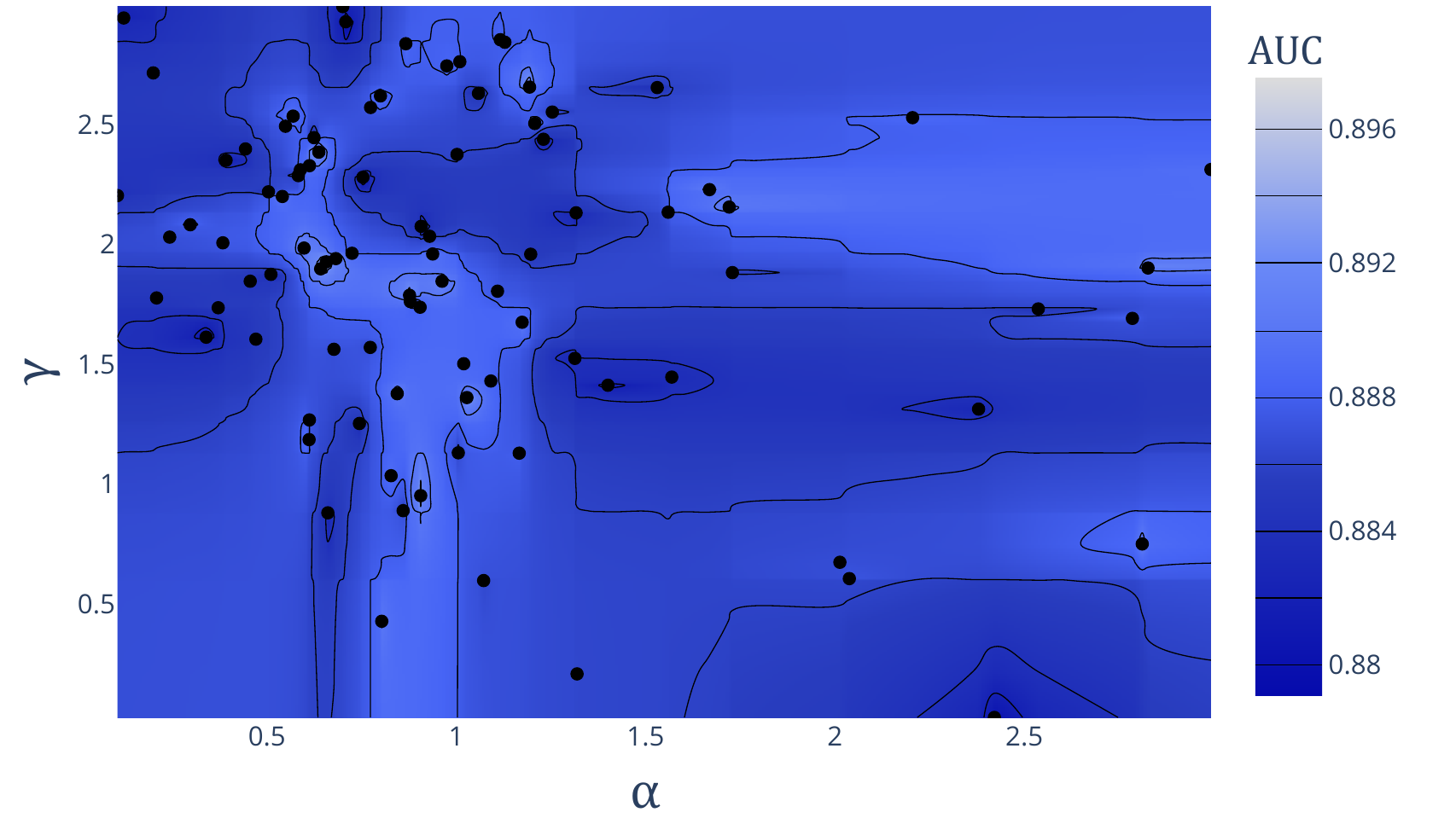}}\\
    \subfloat[][Bitcoin-OTC]{\includegraphics[width=0.8\linewidth]{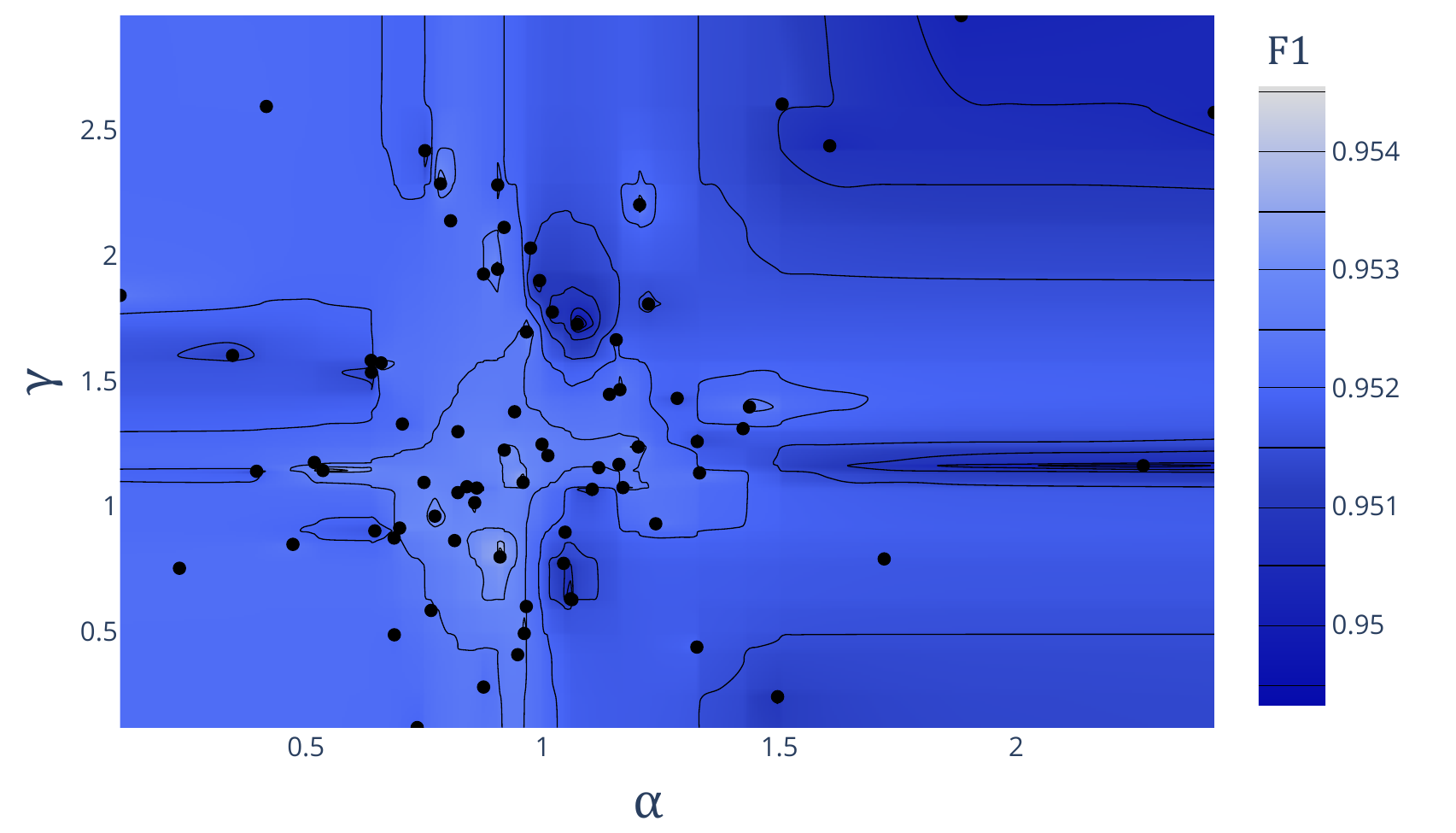}}
    \caption{\color{black}The contour plot of parameter study of loss coefficients $\alpha$ and $\gamma$.}
    \label{fig:contour}
\end{figure}

\noindent\textit{\textbf{Effect of Loss Coefficients.}} To study the effect of the loss coefficients, we conducted the experiments on the proposed SIHG with the varying values of $\alpha$ and $\gamma$. The remaining coefficient $\beta$ is empirically set to 0.83. The search of the optimal loss coefficients is implemented with the optuna~\cite{optuna} toolbox. The Figure \ref{fig:params} is plotted based on the results from 100 trails, with $\alpha$ and $\gamma$ ranging from 0 to 3. {\color{black} In addition, the 2D contour maps of parameter sensitivity evaluated on two datasets are provided in Figure \ref{fig:contour}.} It can be observed that the AUC and F1 scores are going uphill when $\gamma$ increases, which validates the importance of the derived mutual information maximization strategy. Another finding is that, the AUC and F1 scores become quite stable when reaching sufficiently large loss coefficients. This indicates that our \textbf{SIHG} framework is robust with respect to loss coefficients.
\smallskip

\begin{table}[t]
\centering % Centres the table on the page, comment out to left-justify
\caption{Prediction performance of the signed edges with respect to various graph depth on the two datasets.}
\resizebox{1\linewidth}{!}{% 
\begin{tabular}{l c c c c c c c c} 
\toprule % Top horizontal line
% & \multicolumn{4}{c}{\textbf{Bitcoin-Alpha}} & \multicolumn{4}{c}{\textbf{Bitcoin-OTC}} \\ 
% \cmidrule(l){2-5}\cmidrule(l){6-9} 
Dataset &Method &AUC &F1 &macro-F1 &micro-F1 \\ 
\midrule % In-table horizontal line
\multirow{4}{*}{Bitcoin-Alpha} &SIHG-1 &0.8897 &0.9597 &0.6712 &0.9244 \\ 
&SIHG-2 &0.8862 &0.9612 &0.7079 &0.9276 \\
&SIHG-3 &\textbf{0.8981} &\textbf{0.9614} &0.7115 &\textbf{0.9279} \\
&SIHG-4 &0.8664 &0.9602 &\textbf{0.8978} &0.9255 \\ 
\midrule
\midrule
\multirow{4}{*}{Bitcoin-OTC} &SIHG-1 &0.9113 &0.9505 &0.7779  &0.9121 \\
&SIHG-2 &0.9120 &0.9505  &0.7834 &0.9123\\
&SIHG-3 &\textbf{0.9154} &\textbf{0.9528}  &\textbf{0.7949} &\textbf{0.9165}\\
&SIHG-4 &0.9025 &0.9515  &0.7891 &0.9142\\
\bottomrule % Bottom horizontal line
\end{tabular}
}\vspace{-3ex}
\label{tab:depth} 
\end{table} 

\noindent\textit{\textbf{Effect of Graph Depth.}} To investigate the impact of the number of stacked graph layers, we examined the proposed SIHG framework with varying depth of graph layers, \textit{i.e.}, $L\in \{1,2,3,4\}$. As shown in Table \ref{tab:depth}, the performance of the proposed SIHG with deeper graph networks generally increases until $L$ reaches 3. Due to the sparsity of signed networks and the intrinsic over-smoothing risk of GNNs, the SIHG-4 achieves a relatively lower performance compared to SIHG-3.
\smallskip

\noindent\textit{\textbf{Effect of Embedding Dimensionality.}} \textcolor{black}{Another fundamental hyperparameter is the size of the resulting embeddings. We studied the performance of \textbf{SIHG} and the compared baseline \textbf{SGCN} \cite{SGCN} with respect to the dimensionality $d\in \{32, 64, 128\}$ on the Bitcoin-Alpha and Bitcoin-OTC datasets. The results, shown in Table \ref{tab:embed_size} demonstrates the high-dimensional vectors are capable of preserving more information of original network and user interactions, yielding a higher performance for both methods. It is also observed that the proposed method (i.e., SIHG$^*$-32) can learn compact node representations which performance surpasses the one of the longer vector learned by SGCN (i.e., SGCN$^*$-128) for signed link prediction.  }

% \begin{table}[!htb] 
% \centering % Centres the table on the page, comment out to left-justify
% \caption{Recognition accuracies (\%) of the signed edge with respect to various hyerbolic models on the three datasets.}
% \resizebox{1\linewidth}{!}{% 
% \begin{tabular}{l c c c c c c c c} 
% \toprule % Top horizontal line
% & \multicolumn{4}{c}{\textbf{Bitcoin-Alpha}} & \multicolumn{4}{c}{\textbf{Bitcoin-OTC}} \\ 
% \cmidrule(l){2-5}\cmidrule(l){6-9} 
% \textbf{Method} &AUC &F1 &macro-F1 &micro-F1 &AUC &F1 &macro-F1 &micro-F1\\ 
% \midrule % In-table horizontal line

% Euclidean &0.8880 &0.9597 &0.6934 &0.9247 &0.9115 &0.9507 &0.7822 &0.9125\\ 
% Poincare &0.8860 &0.9600 &0.6894 &0.9251 &0.9071 &0.9510 &0.7826 &0.9130\\
% Hyperboloid &\textbf{0.8981} &\textbf{0.9614} &\textbf{0.7115} &\textbf{0.9279}
% &\textbf{0.9154} &\textbf{0.9528}  &\textbf{0.7949} &\textbf{0.9165}\\
% \bottomrule % Bottom horizontal line
% \end{tabular}
% }
% \label{tab:model} 
% \end{table}   

\begin{figure}[t]\vspace{-2ex}
    \centering
    \subfloat[][SGCN]{\includegraphics[width=0.5\linewidth]{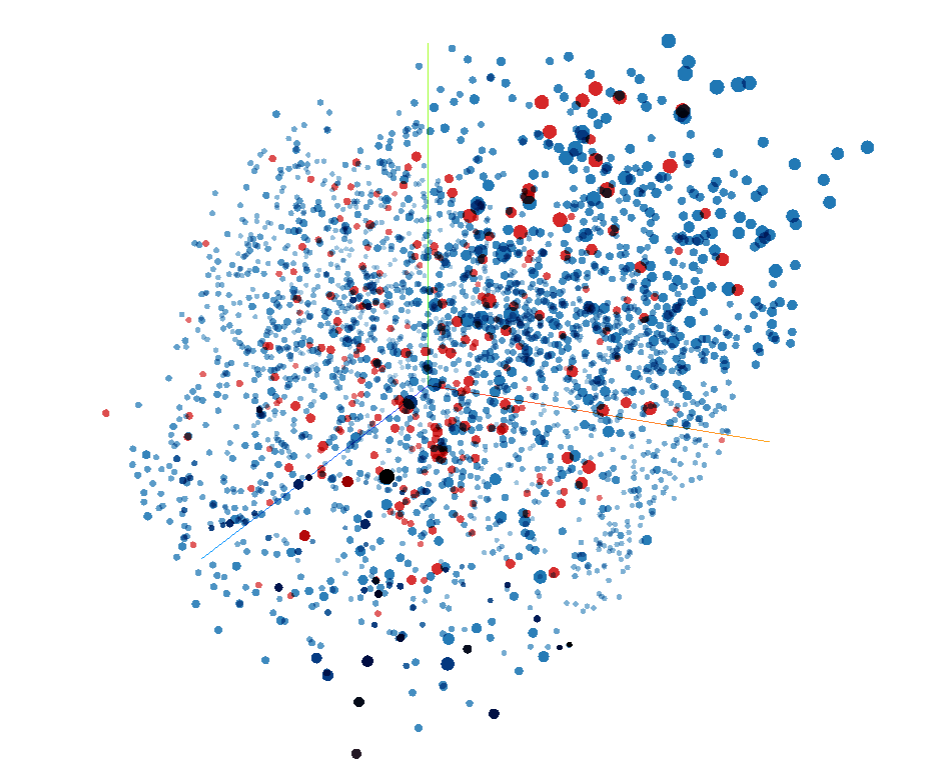}}
    \subfloat[][SiGAT]{\includegraphics[width=0.5\linewidth]{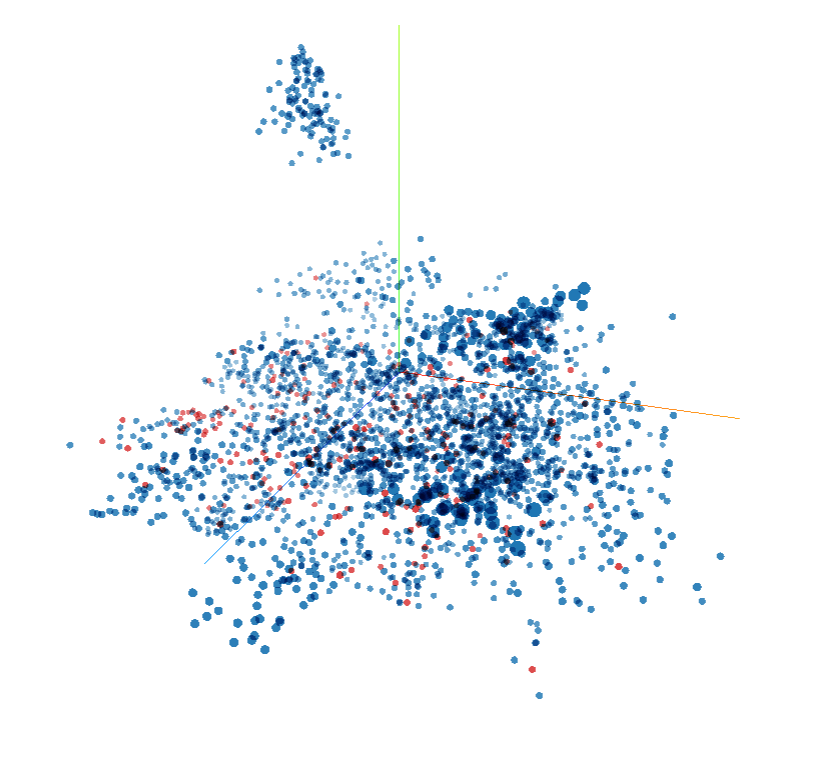}}\\
    \subfloat[][SNEA]{\includegraphics[width=0.5\linewidth]{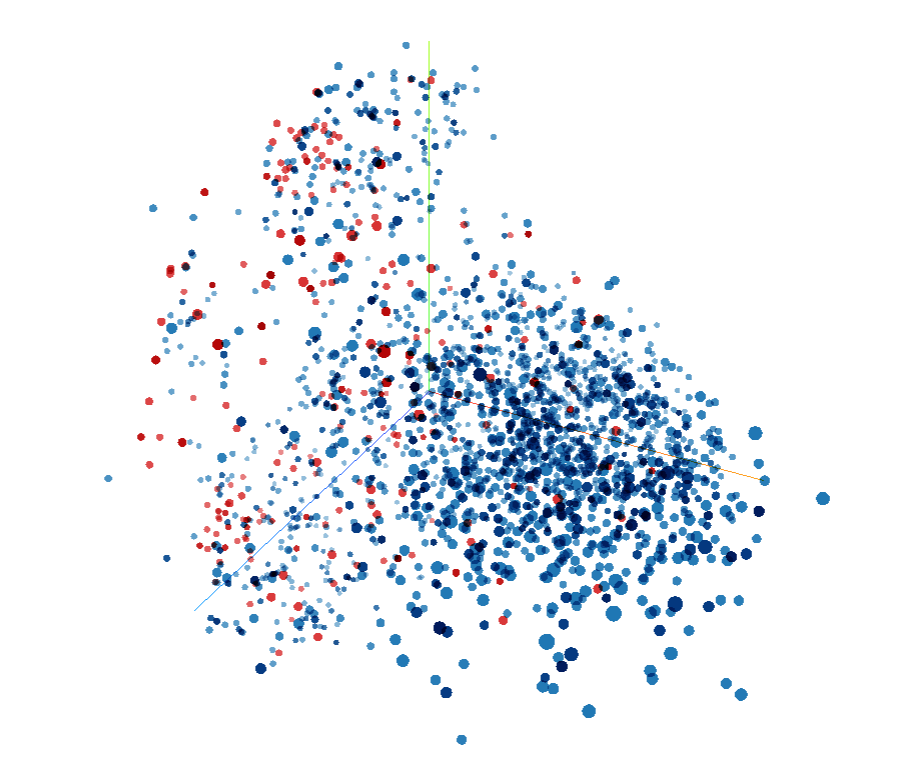}}
    \subfloat[][SIHG]{\includegraphics[width=0.5\linewidth]{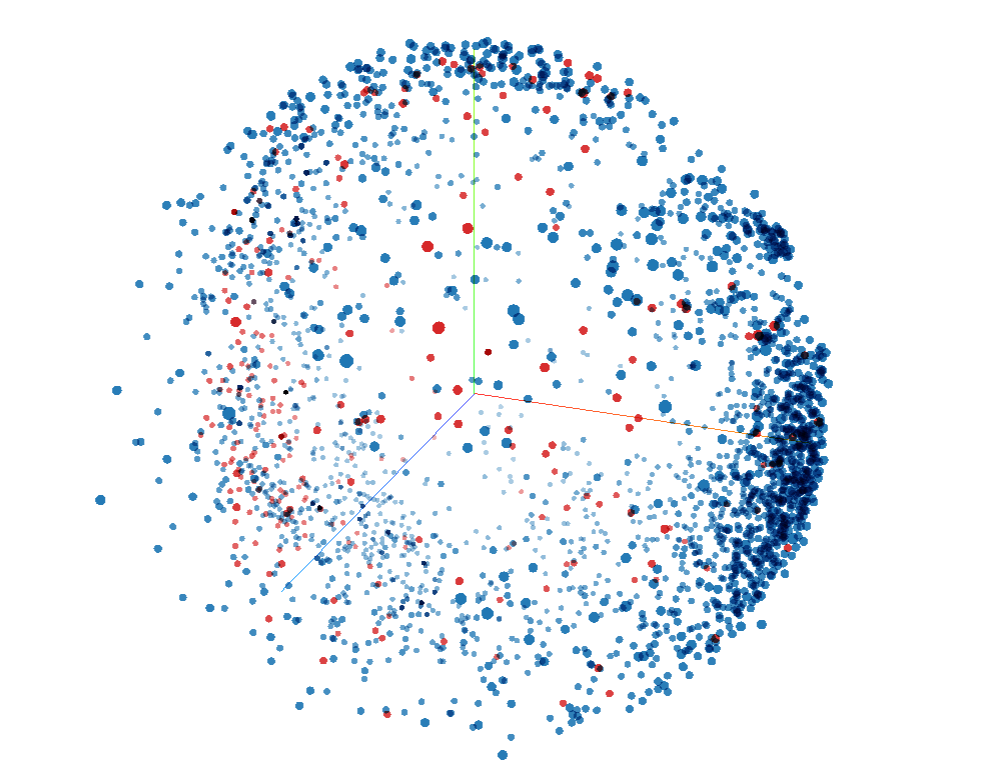}}
    \caption{The visualization of the learned embeddings by graph-based baselines and the proposed SIHG with Hyperbolid model on the Bitcoin-Alpha dataset.}
    \label{fig:pca}\vspace{-3ex}
\end{figure}

\begin{figure}[t]
    \centering
    \includegraphics[width=0.5\linewidth,left]{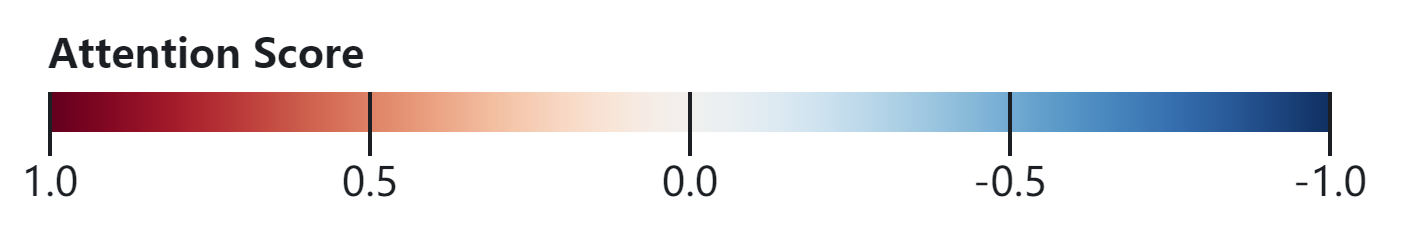}\\\vspace{-3ex}
    % \subfloat[][$w_{ik}^{-(2)}$]{\includegraphics[width=0.5\linewidth]{images/w_ik_neg.pdf}}
    % \subfloat[][$w_{ij}^{-(2)}$]{\includegraphics[width=0.5\linewidth]{images/w_ij_neg.pdf}}
    \subfloat[][$w_{ik}^{-(2)}$]{\includegraphics[width=0.5\linewidth]{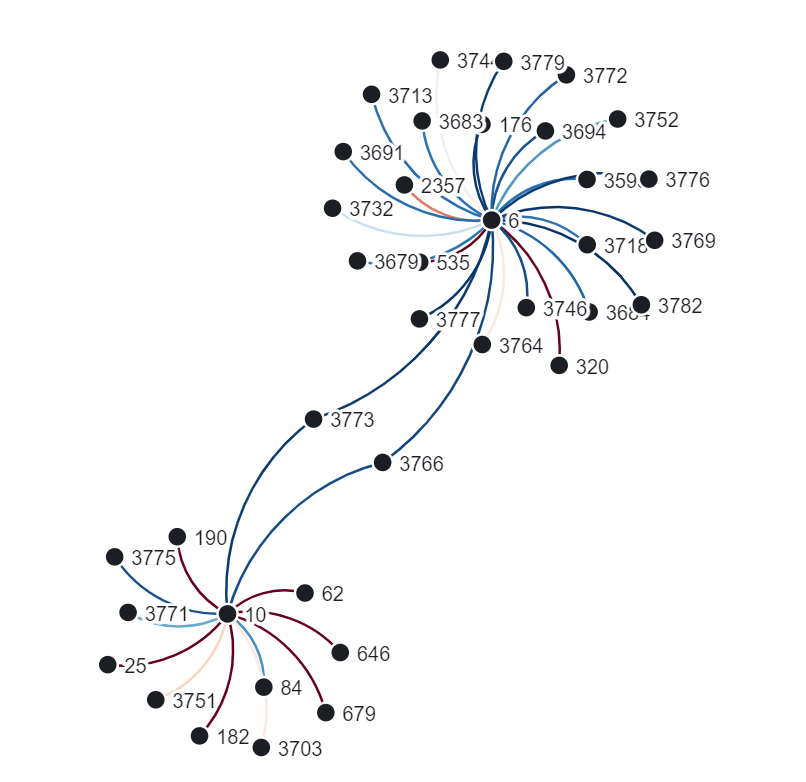}}
    \subfloat[][$w_{ij}^{-(2)}$]{\includegraphics[width=0.5\linewidth]{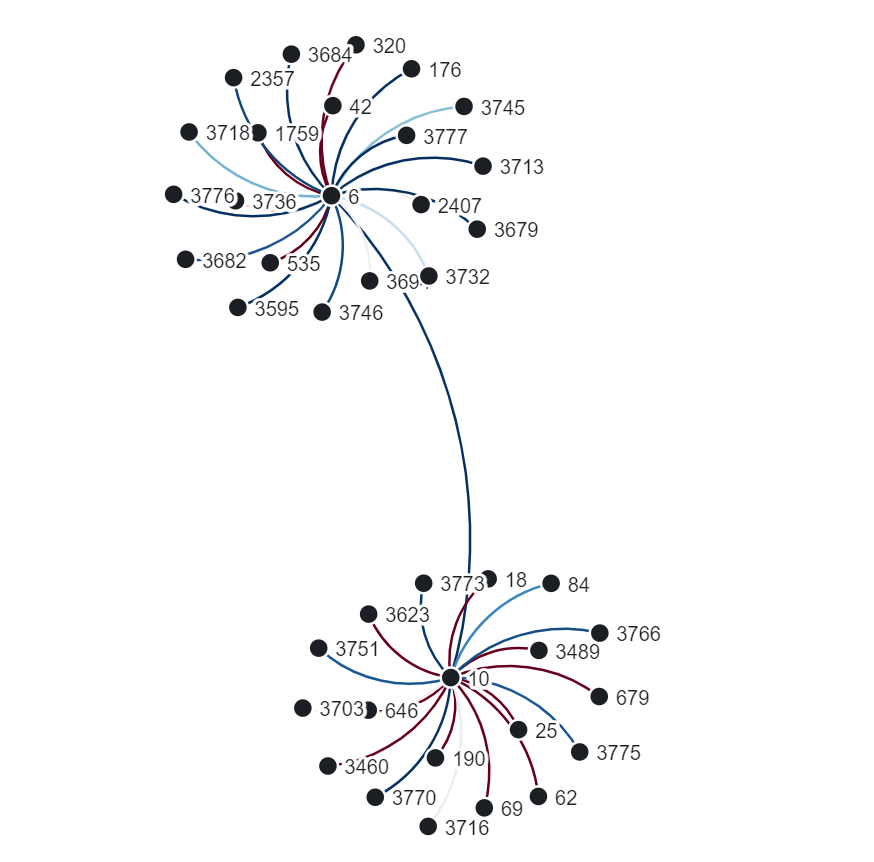}}
    \caption{\color{black}The visualization of the signed attention map extracted from the proposed SIHG. Best viewed in color.}
    \label{fig:att}\vspace{-2ex}
\end{figure}

% \begin{table}[!htb] 
% \centering % Centres the table on the page, comment out to left-justify
% \caption{Recognition accuracies (\%) of the signed edge with respect to various graph depth on the three datasets.}
% \resizebox{1\linewidth}{!}{% 
% \begin{tabular}{l c c c c c c c c} 
% \toprule % Top horizontal line
% & \multicolumn{4}{c}{\textbf{Bitcoin-Alpha}} & \multicolumn{4}{c}{\textbf{Bitcoin-OTC}} \\ 
% \cmidrule(l){2-5}\cmidrule(l){6-9} 
% \textbf{Method} &AUC &F1 &macro-F1 &micro-F1 &AUC &F1 &macro-F1 &micro-F1\\ 
% \midrule % In-table horizontal line

% SHIG-1 &0.8897 &0.9597 &0.6712 &0.9244 &0.9113 &0.9505 &0.7779  &0.9121 \\ 
% SHIG-2 &0.8862 &0.9612 &0.7079 &0.9276 &0.9120 &0.9505  &0.7834 &0.9123\\
% SHIG-3 &\textbf{0.8981} &\textbf{0.9614} &0.7115 &\textbf{0.9279} &\textbf{0.9154} &\textbf{0.9528}  &\textbf{0.7949} &\textbf{0.9165}\\
% SHIG-4 &0.8664 &0.9602 &\textbf{0.8978} &0.9255 &0.9025 &0.9515  &0.7891 &0.9142\\ 
% \bottomrule % Bottom horizontal line
% \end{tabular}
% }
% \label{tab:depth} 
% \end{table}   

\subsection{Embedding Visualization (RQ4)}
To evaluate the importance of high-order topology and hierarchy preservation, we conduct experiments on the test set of the Bitcoin-Alpha dataset, and visualize the node-pair embeddings with PCA in Figure \ref{fig:pca}. The features of the compared baselines, \textit{i.e.}, \textbf{SGCN}, \textbf{SiGAT}, \textbf{SNEA} are extracted before passing them to the classifier. The embeddings are scattered as circles in the projected 3D subspace, with different colors indicating the edge polarities \textit{i.e.}, red represents friendship and blue represents antagonism. In contrast to existing graph models, the proposed \textbf{SIHG} are more powerful to preserve tree structure, where nodes close to the center are generally higher in the hierarchy of the tree. 

\subsection{Interpretability for Social Theories (RQ5)}
In order to study the interpretability of two fundamental social theories, we run experiments on the Bitcoin-Alpha dataset. We extract the signed attention maps of $w^{-(l)}_{ik}$ and  $w^{-(l)}_{ij}$ in the positive and negative aggregation branches at the second graph layer ($l=2$), respectively. \color{black}For simplicity, we only visualize the connections starting from the node $6$ and $10$. The attention maps are plotted in Figure \ref{fig:att} with D3.js. Recalling the aggregation paths in Figure \ref{fig:theory}, we can clearly discriminate the social rules by comparing the sign of attention score for each edge. For instance, the negative edge attention score (e.g., $6\rightarrow 3713$) in $w_{ij}^{-(2)}$ stands for structural balance theory, and the positive one (e.g., $6 \rightarrow 42$) for status theory. The signed attention learned is crucial to provide interpretation of the sociological mechanisms behind the given signed networks.\color{black}

\section{Related Work}
\subsection{Signed Network Embedding}
% As the rapid proliferation of social media data online, there is an urgent need to study complex relationships and interactions among users, which can be modeled as the link prediction task in signed social networks. 
\textcolor{black}{Network embedding approaches~\cite{unsigned,unsigned1,unsigned2,unsigned3,unsigned4,unsigned5,unsigned6} aim to represent each node as a low-dimensional vector, by considering the node's neighborhood and feature information. Similar nodes are expected to be projected close to each other in the subspace, which facilitate the downstream tasks such as node classification and link prediction. Early methods such as DeepWalk \cite{DBLP:conf/kdd/PerozziAS14} and Node2vec \cite{DBLP:conf/kdd/GroverL16} have been proposed to capture node proximity with the random walk strategy. A newly emerging stream of work attempts to compute embeddings of large graph-structured data not in Euclidean but in hyperbolic space, which refers to the space with constant negative curvature. Nickel \textit{et al}. \cite{DBLP:conf/nips/NickelK17} explored the Poincare ball model for embedding learning, which is based on Riemannian optimization. However, prior work only consider learning embeddings for \textit{unsigned networks} that only consist of positive links, thus failing to handle \textit{signed networks} that further consider negative links with more valuable information~\cite{jure}.} The root of modeling signed networks lies in two important sociological theories~\cite{slashdot}, \textit{i.e.}, balance theory and status theory. Motivated by social balance theory, Chiang \textit{et al}.~\cite{clustering} extended weighted kernel k-means clustering to the signed network setting, by considering a signed variant of Laplacian matrix, which can be used as the basis
for graph kernels. Similarly, Zheng \textit{et al}.~\cite{SSE} applied random walk normalized to analyze
signed graphs, which can be embedded in a lower-dimensional
space that reveals the global similarity between nodes. Hsieh \textit{et al}.~\cite{MF} reformulated the sign inference problem as a low-rank matrix completion problem and proved that the missing links can be recovered under certain conditions.

Different from the above-mentioned works that learn node representations by spectral analysis or matrix factorization, another line of work jointly aggregates and propagates information in neural networks. SNE~\cite{SNE} optimizes a Skip-Gram like objective function by the maximum likelihood estimation and incorporates two signed vectors to represent the positive or negative edges with a log-bilinear model. Guided by the extended balance theory, SiNE~\cite{SiNE} introduces a new objective function for signed network embedding, adding virtue nodes to enhance the training process. To improve algorithmic efficiency, SIDE~\cite{SIDE} is built upon a truncated random walk, which aims to represents proximity in signed directed networks as a compact low-dimensional vector. To maintain structural balance in higher-order neighborhoods, SIGNet~\cite{SIGNet} leverages a new targeted node sampling strategy for random walks in directed signed networks. Of late, SGCN~\cite{SGCN} is proposed, which generalizes GCN~\cite{GCN} to signed networks and applies a mean pooling strategy to aggregate messages from neighboring nodes according to balance theory. With the advent in the self-attention mechanism, SiGAT~\cite{SiGAT} utilizes the graph attention networks (GAT) to embed different motifs in directed signed networks. \textcolor{black}{Subsequently, SDGNN \cite{DBLP:journals/corr/abs-2101-02390} extends SiGAT with two additional loss, i.e., margin loss for edge direction and binary cross entropy loss for triangle relation preserving.} Similarly, SNEA~\cite{SNEA} proposes a graph attention layer and provides a more universal way to aggregate information through both positive and negative links based on balance theory. \textcolor{black}{To cope with the over-smoothing problem, SGDNet~\cite{DBLP:journals/corr/abs-2012-14191} derives a signed random walk diffusion method, which aggregates node features on signed edges and effectively exploits information from multi-hop neighbors. 
}
Nonetheless, the existing approaches biasedly rely on the balance and/or status theory for edge sign prediction, which may be easily violated in practice. In our proposed SIHG framework, the principle of mutual information maximization guides the model to infer edge polarities from the informative positive and negative node neighbors, where the aggregation paths are learned by the signed attention module and two social theories are thus naturally unified.

\vspace{-2ex}
\subsection{Mutual Information Estimation}
With the strong growth of data \cite{DBLP:conf/ijcai/ZhangLFY0020,9186366,DBLP:journals/corr/abs-2101-01104,DBLP:conf/cikm/LuoWHYZ18}, Mutual Information (MI) estimation, quantifying the amount of shared information between a pair of random variables, has been playing a pivotal role in representation learning~\cite{infomax} and wide applications~\cite{MIDA}. MI maximization can be used to extract representations that are highly relevant to the target task, or controlling the amount of information between the learned representations and the original data ~\cite{variationalbottle,infobottle,variationalmi}. While effective, few of the prior mutual information estimators can generalize to deep neural networks due to the high dimensionality and sample size. In order to overcome the intractability of MI in the presence of high-dimensional and continuous data, Mutual Information Neural Estimation (MINE)~\cite{mine} makes the estimation of MI on deep neural networks feasible via training a statistics network to distinguish samples coming from the joint distribution and the product of marginals of two random variables. Different from MINE that employs a lower-bound to the MI based on the Donsker-Varadhan representation~\cite{DV} of the KL-divergence, the Jensen-Shannon MI estimator (JSD)~\cite{jsd} follows the formulation of f-GAN KL-divergence. Our proposed SIHG shares the same spirit with the mutual information estimators, which aims to mine the informative representations oriented by the task. But instead of using for unsupervised learning, we, \textit{for the first time}, adapt the mutual information to guide the aggregation in the deep graph model and validate its effectiveness on a practical signed link prediction task.
% Note that the IEEE does not put floats in the very first column
% - or typically anywhere on the first page for that matter. Also,
% in-text middle ("here") positioning is typically not used, but it
% is allowed and encouraged for Computer Society conferences (but
% not Computer Society journals). Most IEEE journals/conferences use
% top floats exclusively. 
% Note that, LaTeX2e, unlike IEEE journals/conferences, places
% footnotes above bottom floats. This can be corrected via the
% \fnbelowfloat command of the stfloats package.
\vspace{-2ex}

\section{Conclusion}
In this work, we propose a deep SIHG framework for the signed link prediction in the presence of large-scale signed social networks. Different from the existing approaches which rely on balance or status theory, we automatically select the aggregation path and reconcile the two theories by maximizing the mutual information between the learned node embeddings and the edge polarities. Experiments evidence the effectiveness of our proposed approach over the state-of-the-art methods, especially improving the AUC scores by up to 13.0$\%$.
\vspace{-2ex}
\ifCLASSOPTIONcompsoc
  \section*{Acknowledgments}
\else
  \section*{Acknowledgment}
\fi

This work is partially supported by ARC FT130101530, NSFC No. 61628206 and Google PhD Fellowship. Thanks to Kevin Swersky for valuable discussions on this topic and to the reviewers for their helpful suggestions.

\ifCLASSOPTIONcaptionsoff
  \newpage
\fi



%\bibliographystyle{IEEEtran}
\vspace{-2ex}
\bibliographystyle{IEEEtran}
\bibliography{IEEEabrv,main}


\vspace{-2ex}
\vspace{-7ex}
\begin{IEEEbiography}[{\includegraphics[width=1in,height=1.25in,clip,keepaspectratio,keepaspectratio]{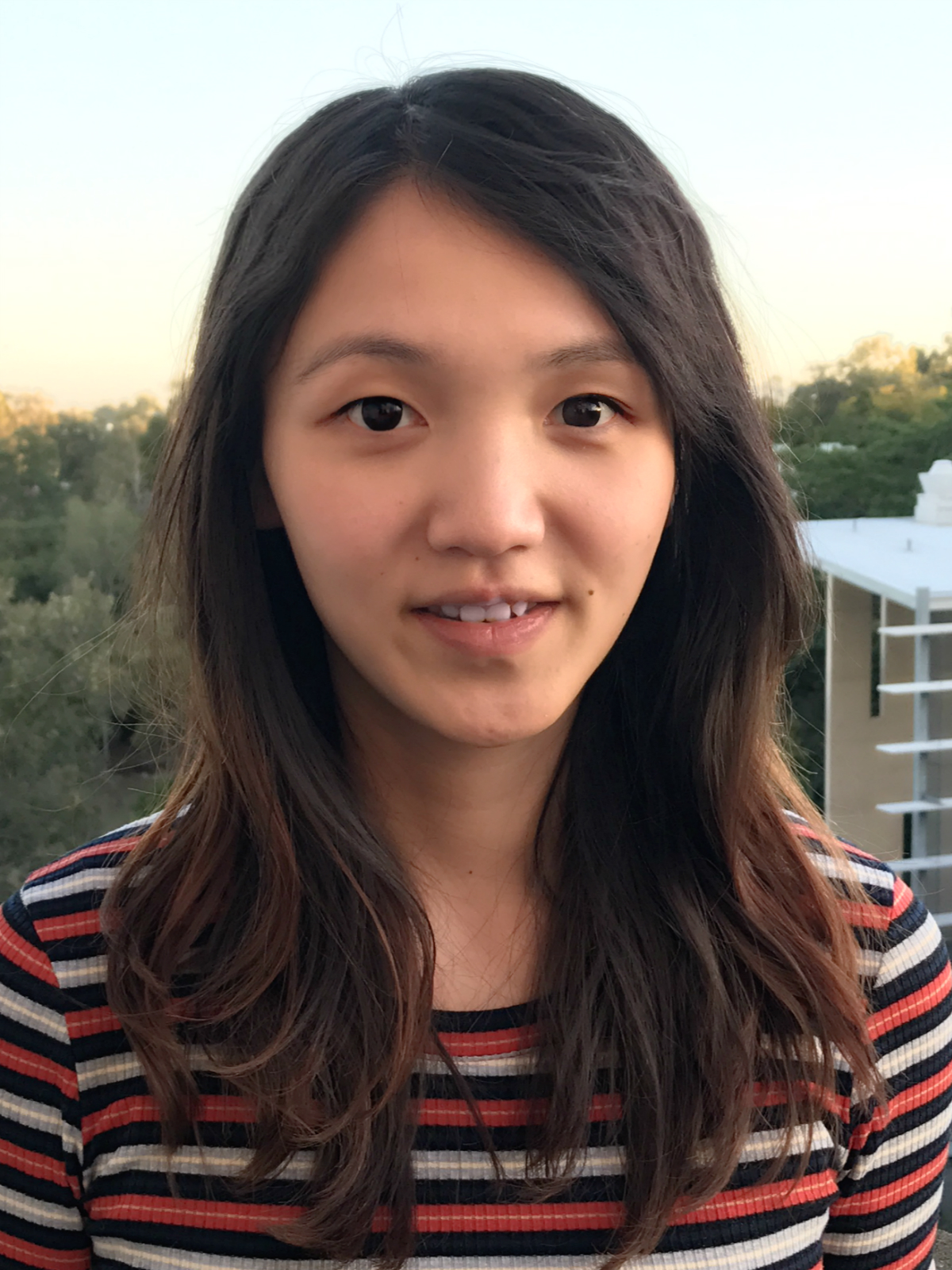}}]{Yadan Luo}
	received the B.S. degree in computer science from the University of Electronic Engineering and Technology of China in 2017. She is currently working toward the Ph.D. degree at the University of Queensland. Her research interests include multimedia retrieval, machine learning and computer vision.
\end{IEEEbiography}
\vspace{-3ex}
\begin{IEEEbiography}[{\includegraphics[width=1in,height=1.25in,clip,keepaspectratio]{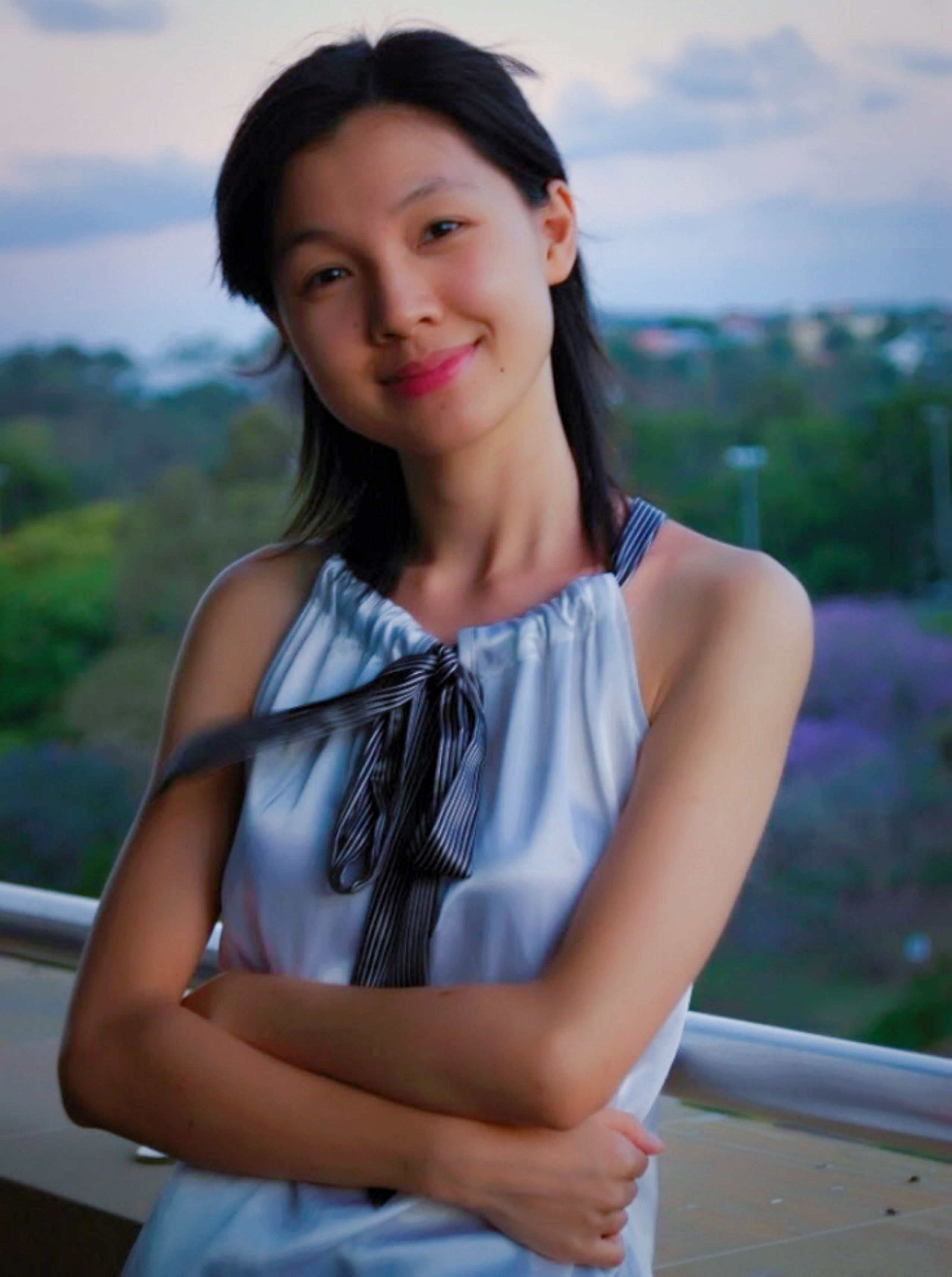}}]{Zi Huang}
	is an ARC Future Fellow in School of ITEE, The University
	of Queensland. She received her BSc degree from Department of Computer Science, Tsinghua University, China, and her PhD in Computer Science
	from School of ITEE, The University of Queensland. Dr. Huang's research
	interests mainly include multimedia indexing and search, social data analysis and knowledge discovery.
\end{IEEEbiography}
% \vspace{-3ex}
% \begin{IEEEbiography}[{\includegraphics[width=1in,height=1.25in,clip,keepaspectratio]{photo/kevin.jpg}}]{Kevin Swersky} is now working as a Research Scientist in Google Brain. Kevin Swersky received his PhD degree in 2016 from The University of Toronto, Canada. He has a broad set of interests in machine learning.
% \end{IEEEbiography}
\vspace{-3ex}
\begin{IEEEbiography}[{\includegraphics[width=1in,height=1.25in,clip,keepaspectratio]{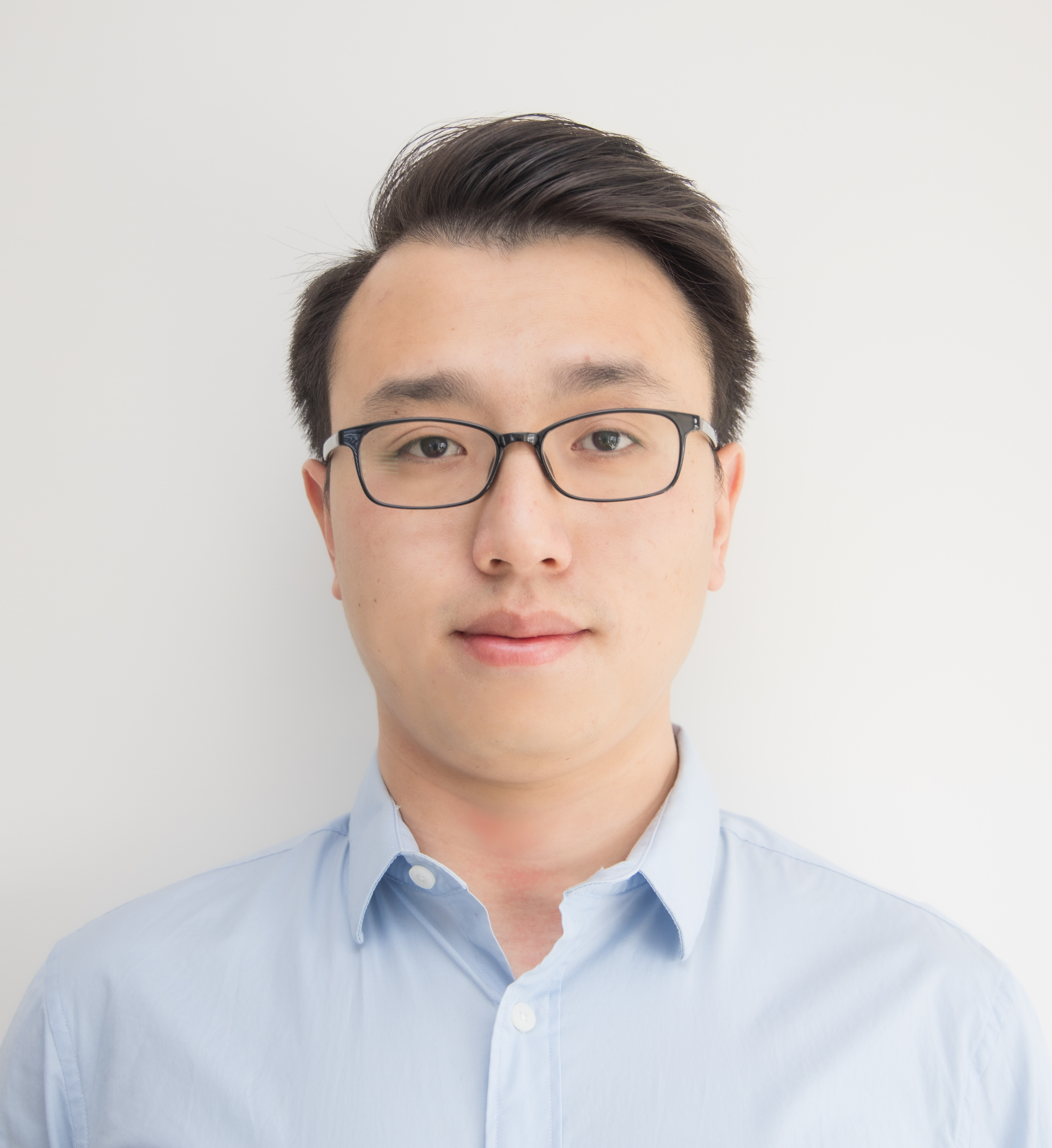}}]{Hongxu Chen} Dr. Hongxu Chen is now working as a Postdoctoral Research Fellow in Network Science Lab at University of Technology Sydney (UTS). Hongxu Chen received his PhD degree in 2020 from The University of Queensland (UQ), Australia. His research interests include data mining, network science, network/graph embedding, recommender systems as well as social networks modelling and analytics.
\end{IEEEbiography}
\vspace{-3ex}
\begin{IEEEbiography}[{\includegraphics[width=1in,height=1.25in,clip,keepaspectratio]{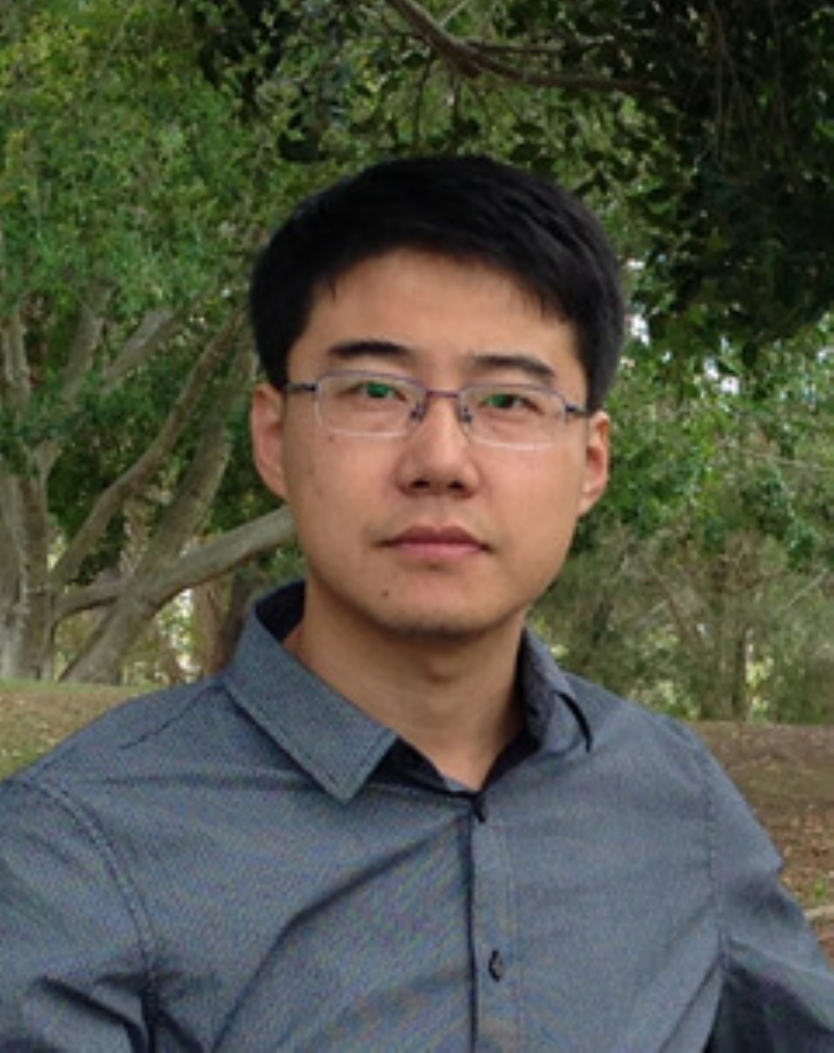}}]{Yang Yang} received the bachelor’s degree from
	Jilin University in 2006, the master’s degree from Peking University in 2009, and the Ph.D. degree from The University of Queensland, Australia,
	in 2012, under the supervision of Prof. H. T. Shen and Prof. X. Zhou. He was a Research Fellow under the supervision of Prof. T.-S. Chua with the National University of Singapore from 2012 to 2014. He is currently with the University of Electronic Science and Technology of China.
\end{IEEEbiography}
\vspace{-3ex}
\begin{IEEEbiography}[{\includegraphics[width=1in,height=1.25in,clip,keepaspectratio]{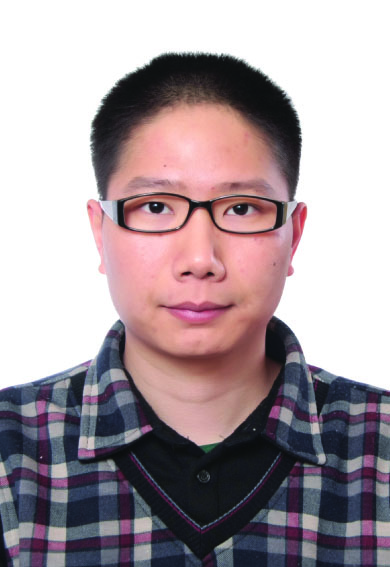}}]{Hongzhi Yin}
received the PhD degree in computer science from Peking University, in 2014. He is a senior lecturer with the University of Queensland. He received the Australia Research Council Discovery Early-Career Researcher Award, in 2015. His research interests include recommendation system, user profiling, topic models, deep learning, social media mining, and location-based services.
\end{IEEEbiography}
\vspace{-3ex}
\begin{IEEEbiography}[{\includegraphics[width=1in,height=1.25in,clip,keepaspectratio]{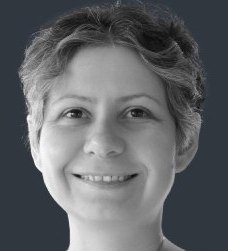}}]{Mahsa Baktashmotlagh} received the bachelor’s degree in software engineering from the Sharif University of Technology and the M.Sc. degree in IT engineering from Polytechnic University in 2007. She also received the PhD degree in computer science from the University of Queensland in 2014. She is now a lecturer in the University of Queensland. 
\end{IEEEbiography}


%\vfill


\end{document}